\documentclass[10pt,conference,letter]{IEEEtran}
\usepackage{epsfig}
\usepackage{graphicx}
\usepackage{epstopdf}
\usepackage{amssymb}
\usepackage{amsmath}
\usepackage{amsthm}
\usepackage{mathrsfs}
\usepackage{algorithm}
\usepackage{algorithmic}
\usepackage{setspace}
\usepackage{enumerate}
\newtheorem{theorem}{Theorem}
\newtheorem{theorem1}{Proposition}
\newtheorem{theorem_def}{Definition}

\begin{document}

\title{Algorithms for Nash and Pareto Equilibria for Resource Allocation in Multiple Femtocells}

\author{
\IEEEauthorblockN{V. Udaya Sankar and Vinod Sharma}\\
\IEEEauthorblockA{Department of Electrical Communication Engineering,\\
Indian Institute of Science, Bangalore 560012, India\\
Email: \{uday,vinod\}@ece.iisc.ernet.in}
}
% avoiding spaces at the end of the author lines is not a problem with
% conference papers because we don't use \thanks or \IEEEmembership
\maketitle

\begin{abstract}
We consider a cellular system with multiple Femtocells operating in a Macrocell. They are sharing a set of communication channels. Each Femtocell has multiple users requiring certain minimum rate guarantees. Each channel has a peak power constraint to limit interference to the Macro Base Station (BS). We formulate the problem of channel allocation and power control at the Femtocells as a noncooperative Game. We develop decentralized algorithms to obtain a Coarse Correlated equilibrium that satisfies the QoS of each user. If the QoS of all the users cannot be satisfied, then we obtain a fair equilibrium. Finally we also provide a decentralized algorithm to reach a Pareto and a Nash Bargaining solution which has a much lower complexity than the algorithm to compute the NE.
\end{abstract}
\emph{\maketitle{Keywords-}}
Femtocell, QoS, game theory, decentralized algorithms, Nash Bargaining.

\section{Introduction}
Currently a significant fraction (80\%) of traffic in the cellular systems is being generated indoors. Out of this about $60\%$ is voice traffic which requires a good quality of the received voice. However, due to attenuations from the walls, the quality of this voice will not be good unless the Base station (BS) transmits at power levels which are no longer legally allowed in many countries. Hence, femtocells (\cite{vick1}, \cite{ghosh}) are being considered as an option.

A Femtocell (FC) is a small cell meant to cater to the wireless users within a building. It consists of a Femtocell Access Point (FAP), which is a low powered, small, inexpensive base station deployed inside the  building. It is connected to a wireline network. If a Mobile Station (MS) is within the building and encounters an outage from the main BS in the cell (called Macro BS), it can be handed over to the FAP deployed within the building. This improves the quality of service (QoS) to the indoor applications and also offloads a significant fraction of cellular traffic to the wireline network.

Although, deployment of FAPs in a macrocell (MC) environment improves the performance of the users inside the FC, it causes interference to the MSs in the MC and adjacent FCs. Therefore, to reap the benefits of a FC one needs careful interference management (\cite{zahir}). An upper limit must also be imposed on the transmit power of the MSs within a FC. Reducing the transit power in a FC will also help reduce the carbon footprint of the cellular systems (\cite{Hasan}, \cite{shakir}). We address this problem in this paper.

In the following we provide the related literature survey. A good overview on the topic is provided in {\cite {vick1}, \cite{chow}, \cite{rao}}. The problem of power control and channel allocation among FCs is addressed in {\cite{abdel}, {\cite{ref_b}}}. In {\cite{abdel}} the authors try to admit as many users as possible while in {{\cite{ref_b}}} QoS of delay sensitive users is taken care of by ensuring them minimum rates.\vspace{-0.074cm}

A dynamic resource management scheme for LTE based hybrid access FC is proposed in \cite{yloong}. Multi-objective resource allocation for FC networks is proposed in \cite{yllee}.  In \cite{jxu}, the authors have discussed cooperative distributed radio resource management algorithms for time synchronisation, carrier selection and power control for hyper-dense small cell deployment. In \cite{ref_90}, the authors propose a cognitive radio resource management scheme for FCs to mitigate cross-tier interference.

The problem of power control to different channels within each FC via Game theory is addressed in {\cite{barba}}, {\cite{mehdi}}, {\cite{vikram}}, {\cite{hong}}, {\cite{ref_a}}. In {\cite{barba}} the aggregate throughput of FCs is considered and Nash Equilibria (NE) obtained. In {\cite{mehdi}} a stackelberg game is formulated. In {\cite{vikram}} and {\cite{trong}} a Pareto point is obtained. The problem of power control and channel allocation via game theory is addressed in {\cite{ref_d} and {\cite{uday1}}. This problem is more complex because of mixed integer programming involved. {\cite{ref_d} uses auction theory to do resource allocation while {\cite{uday1}} develops efficient distributed algorithms to obtain NE while guaranteeing minimum rates to individual users. In {\cite{uday1}} the scenario when it is not possible to provide minimum rates to individual users is also addressed by providing 'fair' NE.\vspace{-0.1cm}

In this paper we continue to study the setup in {\cite{uday1}}. As opposed to {\cite{uday1}}, we develop low complexity distributed algorithms that can be used whether the QoS of the users are satisfied or not. Also we develop algorithms for computing Pareto optimal points and Nash bargaining solutions. These were not obtained in {\cite{uday1}}. We also show how our setup can be extended when there are voice and data users in the system. As against {\cite{vikram}} and {\cite{trong}}, we have multiple users in each FC. Thus in {\cite{vikram}} and {\cite{trong}}, each considering one user only in each FC, there is only the power allocation problem while we consider power and channel allocation. Also, {\cite{trong}} does not have any QoS requirements at the FCs. While  {\cite{vikram}} has a single channel and considers an SINR requirement.

This paper is organized as follows. Section II describes the mathematical model of the problem. Section III provides a game theoretic framework and also provides efficient algorithms for a Coarse correlated equilibrium (CCE). Section IV provides an algorithm for a Pareto point and Section V extends it to obtain a Nash Bargaining solution. Section VI considers a system which has voice and data users and we ensure QoS to the voice users. Section VII shows the efficacy of the algorithms via a few examples. Section VIII concludes the paper.
\section{Mathematical Model}
We consider a two tier cellular system in which within a MC there may be many FCs. The cellular system has multiple subchannels, perhaps using OFDMA (e.g., LTE/LTEA) and the subchannels are shared by the FCs and the outdoor users in the MC. The transmission between a FAP and its MSs may be in the uplink or downlink or in both directions. The allocation of channels and powers to different users will be done by the FAP using the same algorithms (although the peak power constraint in the two directions can be different for the same channel). Thus, for simplicity we will consider down link only.\vspace{-0.1cm}

The MSs using a FC are also the MSs for the Macro BS. Thus, these MSs can get the information from the Macro BS about which subchannels are being used by the users in the MC (e.g., from the DL/UL map in Wimax, PDCCH/PUCCH in LTE sent by the MBS in each frame). The indoor MSs can also sense the SINR in each subchannel and can further be directed by the MBS on the maximum power they can use in transmitting in different available subchannels. This information can be sent to the FBS by the MSs within its domain or by the MBS directly (see e.g., {\cite{ref6}). FBS uses this information to decide on subchannel allocation and power control within its FC to provide QoS to its users while using minimum power within the limits prescribed by the MBSs. Minimizing power reduces interference to other FCs and to MSs outside. We assume a control channel via which the different FCs can coordinate with each other and self configure.\vspace{-0.1cm}

Let there be $K$ FCs sharing $N$-subchannels deployed in a MC. Each FC $k$ consists of $M_k$ users which have minimum rate requirement $\bar{R}^k_j, j=1, 2, ..., M_k$. Each channel $i$ has maximum power constraint $\bar{P}^k_i, i=1, 2, ..., N$ in FC $k$. Let $G^k_{i,j}$ be the channel gain of $i^{th}$ channel for $j^{th}$ user in FC $k$. Let the interference from MC for user $j$ in FC $k$ on channel $i$ be $I^k_{i,j}$. We assume that the receiver noise variance $\sigma^2$ is same for all FCs and channels (although this can also be different and our setup will require no change for this). Let $G^{k,l}_{i,j}$ be channel gain at channel $i$ for user $j$ in FC $k$ from FC $l$. Based on this information, each FC $k$ has to compute power $P^k_i$ allocated to each subchannel $i$ and subchannel allocation $A^k_{i,j},$ where

\[A^k_{i,j}=\left \{
\begin{array} {ccl}
1, & &\textrm{if subchannel } i \textrm{ is assigned to user } j, \\
0, & &\textrm{otherwise},
\end{array}
 \right. \]
such that the overall power used in the FC is minimized while the users' QoS is satisfied. Thus each FC $k$ solves the optimization problem,
\begin{equation}
min \quad \sum_{i=1}^N P^k_i
\end{equation}
such that
\vspace{-2mm}
\begin{equation}
       \sum_{i=1}^NC^k_{i,j} A^k_{i,j} \geq \bar{R_j}^k,\quad \forall j=1,2,...,M_k,
\end{equation}
\begin{equation}
       P^k_i\leq \bar{P_i}^k,\quad \forall i=1,2,...,N,
\end{equation}
\begin{equation}
       \sum_{j=1}^{M_k} A^k_{i,j} \leq 1, \quad \forall i=1,2,...,N.
\end{equation}
where $C^k_{i,j} = \log_2\left( 1+\frac{{P^k_i G^k_{i,j}}}{\Gamma(\sigma^2 + I^k_{i,j} + \sum_{l \neq k}G^{l,k}_{i,j}P^l_i)}\right)$ is the transmission rate for user $j$ on channel $i$ in FC $k$ if channel $j$ is allocated to it and $\Gamma$ is the SNR gap included for practical rates achievable depending on the modulation and coding scheme (\cite {chow1}).

Equation (2) specifies the minimum rate requirements, while (3) specifies the power constraints on each channel. The constraint (4) ensures that any subchannel is allocated to only one user within a FC. Also, while trying to provide QoS to each of its users, each FC also wants to minimize the total power (1) it needs, to reduce the costs involved as well as to address the green communication related issues (\cite{Hasan}). Since the decisions made by each FC affect the decisions of other FCs, we address this problem as a game.
\section{game theoretic solution}
\subsection{Game formulation and solution}
We formulate the game for our system as $\mathscr{G} = \langle \mathscr{I}, \mathscr{X}, (\Phi_k(x))_{k \in \mathscr{I}}\rangle$, where $\mathscr{I} = \{1,2,...,K\}$, the set of FCs is the set of players, $\mathscr{X}$ is the overall strategy space and $\Phi_k$ is the utility of player $k$. Let $\underline{P}^k = \{P^k_i,\quad i=1,2,..., N\}$. We define strategy set $\mathscr{X} = \{(\underline{P}^k \in R^N_+, A^k_{i,j} \in \{0,1\}, k=1,2,..., K): \sum_{i=1}^NC^k_{i,j} A^k_{i,j} \geq \bar{R_j}^k, j=1,2,...,M_k, P^k_i \leq \bar{P_i}^k, \sum_{j=1}^{M_k} A^k_{i,j} \leq 1, i=1,2,...,N\}$. We are interested in finding a decentralized energy efficient Nash Equilibrium ({\cite{palomar}}) which provides QoS to each user in the system (if at all possible). If it is not possible we provide a \emph{fair} NE (to be defined later).

Define the utility function for FC $k$ by
\begin{equation}
\Phi_k(\textbf{x}) = - \sum_{i=1}^NP^k_i, \forall \textbf{x} \in \mathscr{X}.
\end{equation}
Maximizing $\Phi_k(\textbf{x})$ will minimize the total power used by FC $k$. It is easy to verify that the game $\mathscr{G}$ defined above is an exact \textit{Potential} game (\cite{shapely}) with Potential function
\begin{equation}
f(\textbf{x}) =  - \sum_{k=1}^K\sum_{i=1}^NP^k_i.
\end{equation}

We observe that $\mathscr{X}$ is a compact set. Also the potential function $f$ is continuous. Therefore, it has a global maximizer
and hence has a (generalized) NE (\cite{Han1}, {\cite{palomar}}). In the following we provide distributed algorithms to compute a NE.

Let
\begin{equation}
\overline{\mathscr{D}}_k(\textbf{x}_{-k}) = \{\textbf{x}^*_k \in \mathscr{X}_k(\textbf{x}_{-k}): \textbf{x}^*_k= \arg \max_{\textbf{x}_k} \Phi_k(\textbf{x}_k, \textbf{x}_{-k})\},
\end{equation}
where $\mathscr{X}_k(\textbf{x}_{-k})$ is the set of strategies of $k$ which are possible when the strategies of all other users are $\textbf{x}_{-k}$. The set $\overline{\mathscr{D}}_k(\textbf{x}_{-k})$ provides strategies for player $k$ which maximize its utility for a fixed strategy $\textbf{x}_{-k}$ by all other users. For a given $\textbf{x}_{-k}$, $\textit{best response}$ provides an element of $\overline{\mathscr{D}}_k(\textbf{x}_{-k})$.

Let for an $\epsilon >0$ and $\textbf{x}_{-k}$,
\begin{equation*}
\mathscr{D}_k(\textbf{x}_{-k}) = \{\textbf{x}^*_k \in \mathscr{X}_k(\textbf{x}_{-k}):  \Phi_k(\textbf{x}^*_k, \textbf{x}_{-k}) \geq \Phi_k(\textbf{x}_k, \textbf{x}_{-k}) + \epsilon,
\end{equation*}
\begin{equation}
\forall \textbf{x}_k\}.
\end{equation}
Then $\epsilon$-$\textit{Better Response}$ provides a point in $\mathscr{D}_k(\textbf{x}_{-k})$.\\
\vspace{-0.01cm}
 For a potential game better response and $\epsilon$-better response based iterated algorithms usually converge. However, for our problem the feasible strategy set of a player, due to (2) is dependent on the strategies being played by the other players. Therefore, for the resulting (generalized) NE the best response dynamics may not converge {\cite{decom}}.\\
\vspace{-0.01cm}
The $\textit{$\epsilon$-Better Response algorithm}$ obtains, at iteration $n+1$, at each FC $k$, $k=1, 2, ..., K$,
\begin{equation}
\textbf{x}^{n+1}_k \in \mathscr{D}_k(\textbf{x}^{n+1}_1, \textbf{x}^{n+1}_2,.., \textbf{x}^{n+1}_{k-1}, \textbf{x}^n_{k+1}, .., \textbf{x}^n_K),
\end{equation}
and then passes on the $\epsilon$-better response to FC $k+1$.\\
\vspace{-0.01cm}
Since the potential function is bounded and in the $\epsilon$-better response algorithm, the potential function increases by atleast
$\epsilon$, it converges in a finite number of steps to an $\epsilon$-Nash Point (\cite{neel}).

For computationally simple algorithms, we consider random-better response (or random $\epsilon$-better response) which converges for the above potential game. These algorithms have lower complexity per iteration.
But for a continuous strategy space one may need a large number of iterations. Thus, we have developed a novel variation of these algorithms in {\cite{uday1}}
which converges much faster. This algorithm assumes that there are enough resources available in the system so that the QoS of each user in each FC can be satisfied.

If the QoS of all the users in all the FCs cannot be satisfied then we have developed a distributed algorithm in {\cite{uday1}} to obtain a ``fair" NE. In this case, we try to satisfy the largest fraction of QoS of all the users in each FC. In particular, for each FC $k$ we obtain power $\underline{P}^k$ and subchannel allocation $A^k$ that
%when $P^k_i = \bar{P}^k_i, \forall i, k$, it is possible to allocate largest fractin of the rates that can be satisfied for all users in cell $k$. Such 'fair' objective functions for a single player case have been studied before ({\cite{kodi}, \cite{vandana}}) and is a max-min criterion for fraction of rates received. Here we consider max-min criterion for fairness rate allocation.

%Define ${C}^k_{i,j} = \log_2\left( 1+\frac{{P^k_i G^k_{i,j}}}{\Gamma(\sigma^2 + I^k_{i,j} + \sum_{q \neq k}G^q_{i,j}A_{i,j}^q\bar{P}^q_i)}\right)$ . Find $A_{i,j}^k, P^k_i,  \forall i, \forall j, k = 1,2,..., K, 0\leq\alpha_k\leq1 $
\begin{equation}
\max \quad \alpha_k
\end{equation}\vspace{-7mm}\\
such that \vspace{-3mm}\\
\begin{equation}
\frac{1}{\bar{R}^k_j}\sum_{i=1}^N{C}^k_{i,j} A_{i,j}^k \geq \alpha_k, \quad \forall j = 1,2,...,M_k,
\end{equation}

For this problem, since $\mathscr{X}_k$ is compact and utility functions $\alpha_k$ are continuous, the game has a (mixed) NE (MNE) {\cite{fuden}}. However, it is not a Potential game. In {\cite{uday1}} we discretized the powers and provided a regret matching algorithm based on  {\cite{hart2}} which converges to a correlated equilibrium (CE) of the discretized game. It can be shown that as the discretization step goes to zero, the CE of the discretized game converges to that of the original game(\cite{noah}, \cite{gills}).

When the QoS of all the users cannot be satisfied in each FC, then $\alpha_k$ of the solution is $ < 1$ at least for some $k$. If it can be satisfied, then we may see that at the solution point $\alpha_k \geq 1$ for all $k$. Thus, this setup can actually handle all the cases.
\subsection{Algorithms}
\subsubsection{Coarse Correlated Equilibrium}
In this paper for the problem (3)-(4), (10)-(11) we present another algorithm, based on multiplicative weights algorithm {\cite{rough}}. This algorithm has less computational complexity than the regret matching algorithm but converges to the set of coarse correlated equilibria (CCE) instead of, to the set of correlated equilibria (CE).
%In CCE every player deviations are depend on distribution of outcome as opposed to distribution of outcome and recommended strategy in CE. This property of CCE leads to easier computation of CCE as compared to CE {\cite{garden}}. Furthermore, its price of anarchy bound in a reasonable large class of games is no worse than that of NE {\cite{garden}}.
\begin{theorem_def}
A distribution $\sigma$ on the set $\mathscr{X} = \prod_{k=1}^K \mathscr{X}_k$ of strategies is a CCE{\cite{rough}} if for every player $k\in \{1, 2, ..., K\}$ and for every unilateral deviation $x'_k \in \mathscr{X}_k$
\begin{equation*}
E_{\textbf{x}\sim \sigma}[\Phi_k(\textbf{x})] \geq E_{\textbf{x}\sim \sigma}[\Phi_k(x'_k, \textbf{x}_{-k})]
\end{equation*}
where $\textbf{x}\sim \sigma$ denotes that the overall strategy $\textbf{x}$ has distribution $\sigma$ and on the right hand side $x_{-k}$ has the marginal distribution from $\sigma$.
\end{theorem_def}
Since MNE $\subseteq$ CE $\subseteq$ CCE, CCE also exists for our problem. Furthermore, the price of anarchy of CCE is shown to be no worse than that of a pure NE for a reasonably large class of games ({\cite{garden}}).
%*********************************

To obtain a CCE for our problem, we again discretize the powers to obtain a finite game. For this finite game, the algorithm to obtain a CCE is provided in Algorithm 1. In this algorithm, we first use the multiplicative weights algorithm (Algorithm 2) to obtain a CCE. If $\alpha_k<1$ for some $k$ then we retain that solution. But we reduce (via KKT) the power allocation of the users which are getting more rate than their requirements, so that the overall powers are reduced and $\alpha_k$ of those users becomes 1. This will not reduce $\alpha_k$ of others users. But it can possibly increase $\alpha_k$ of users with $\alpha_k<1$ because their interference is reduced. If $\alpha_k \geq 1$ for all $k$, then we retain the channel allocation of all the FCs but play a power game among the users to minimize the total sum powers while making $\alpha_k=1$ for each $k$.
\begin{algorithmic}
\begin{algorithm}
\caption{Allocation Algorithm 1}
\STATE $1.$Compute the CCE from Algorithm 2 (Algo-CCE).
\STATE $2.$ If $\alpha_k < 1$ for some $k$ then stop.
\STATE $3.$ if($\alpha_k \geq 1$) for all $k$, then fix $A^k_{i,j}$ for all $(i,j)$ and play power allocation game.
\end{algorithm}
\end{algorithmic}
\begin{algorithmic}
\begin{algorithm}
\caption{Algo-CCE}
\STATE $(a)$ Initialise $W^0_k(x_k) = 1, \forall x_k \in \mathscr{X}_k$ and $0 < \epsilon < \frac{1}{2}$
\STATE $(b)$ $ n \geq 1$
\STATE  \begin{enumerate}
\item  Play an action according to distribution $p^n_k = \frac{W^n_k}{\Gamma^n_k}$, where $\Gamma^n_k = \sum_{x_k \in \mathscr{X}_k}W^n_k(x_k)$
\item  Each player $k$ receives a cost vector $C^n_k$, where $C^n_k(x_k) = E_{\textbf{x}_{-k}\sim p^{n-1}_{-k}}[C_k(x_k, \textbf{x}_{-k})], \forall x_k$, where $p^{n-1}_{-k} = \prod_{j\neq k} p^{n-1}_j$
\item  For given cost vector $C^n_k$, update $ W^{n+1}_k(x_k) = W^n_k(x_k)(1 - \epsilon)^{C^n_k(x_k)}, \forall x_k \in \mathscr{X}_k$
\end{enumerate}
\STATE $(c)$ Repeat (1), (2), (3) until convergence.
\end{algorithm}
\end{algorithmic}

The power allocation game for (1)-(4) for fixed subchannel allocation matrices $A^k$, to be used in Algorithm 1 is a potential game with coupled constraints. Its generalized NE exists because the utility function is convex, continuous and the strategy space is convex (\cite{palomar}). Thus, the best response algorithm (which in this case can be obtained by solving the KKT equations) converges (\cite{palomar}).

To execute Algorithm 2, the FCs have to exchange their mixed strategies with each other. The algorithm works whether all FCs update their strategies simultaneously or sequentially.

Algorithm 2 is written for cost functions between $[0,1]$. Therefore, instead of taking $\alpha_k$ as the utility to maximize, we take $C_k(\alpha_k)=1-2\alpha_k+\alpha^2_k$ as the cost to minimize. If $\alpha_k \leq 2$ then this function has a unique minimum at $\alpha_k=1$, the point we want to obtain. From either side of $1$, this function monotonically decreases towards $1$. Since we are trying to minimize power while satisfying QoS of different users, if $\alpha_k>2$ is possible, we can reduce $\bar{P}^k_i$ of that FC such that $\alpha_k \leq 2$, without sacrificing performance.

Another way to address our problem is via penalty function approach. For $\rho > 0$, consider the cost function \begin{equation}
C_k(x_k,\textbf{x}_{-k}) = \sum_{i=1}^N P^k_i + \rho \sum_{j=1}^{M_k}\max\{0,\bar{R_j}^k-\sum_{i=1}^NC^k_{i,j} A^k_{i,j}\}.
\end{equation}
For this game also, the strategy space $\mathscr{X}_k = \{(\underline{P}^k \in R^N_+, A^k_{i,j} \in \{0,1\}:  0\leq P^k_i\leq\bar{P^k_i}, \sum_{j=1}^{M_k}A_{i,j}^k, \leq 1\quad i=1,2,...,N \}$ is compact and the cost function $C_k(x_k, \textbf{x}_{-k})$ is continuous. Hence the game has (mixed) NE{\cite{fuden}}.

To apply Algorithm 1, we consider the normalised cost
%****************************************************
\begin{equation*}
\alpha_{nor}^k = \frac{1}{\rho+1}\biggl(\frac{1}{N}\sum_{i=1}^N (\frac{P^k_i}{\bar{P_i}^k})+
\end{equation*}
\begin{equation}
\frac{\rho}{M_k}\sum_{j=1}^{M_k}(\frac{\max\{0,\bar{R_j}^k-\sum_{i=1}^NC^k_{i,j} A^k_{i,j}\}}{\bar{R_j}^k})\biggr).
\end{equation}
%\begin{equation}
%+ \frac{\rho}{M_k}\sum_{j=1}^{M_k}(\frac{\max\{0,\bar{R_j}^k-\sum_{i=1}^NC^k_{i,j} A^k_{i,j}\}}{\bar{R_j}^k})\biggr).
%\end{equation}
We will call this Algorithm 2A.

For this problem formulation, we do not need to play a power game after the CCE is obtained because when $\alpha_k \geq 1$ for all $k$, due to the cost function chosen, for reasonable values of $\rho$, the rates obtained for the different users will not be substantially more than $\bar{R}^k_i$. Thus we can use Algorithm 2 only. However, the CCE for (10) will generally be fair (within a FC) when the QoS are not satisfied while for the present formulation it may be less likely.

We will compare the solutions obtained by Algorithm 1 and Algorithm 2A in section VII.
\section{Pareto Optimal Points}
The CCE obtained from the above algorithms (Algorithms 1-2, 2A) may not provide good overall system  performance. Thus, in the following, we also develop a distributed algorithm for computation of a Pareto optimal point.
\begin{theorem_def}(\cite{Lasaulce}, \cite{Miettinen})
A decision vector $\textbf{x}^* \in \mathscr{X}$ is Pareto optimal if there does not exist another decision vector $\textbf{x} \in \mathscr{X}$ such that $\Phi_k(\textbf{x}) \geq \Phi_k(\textbf{x}^*), \forall k=1, 2, ..., K$ and $\Phi_j(\textbf{x}) > \Phi_j(\textbf{x}^*)$ for at least one index $j=1, 2, ..., K$.
\end{theorem_def}
Let, for $\beta_k >0, k=1, 2, ..., K$,
\begin{equation}
\Phi(\textbf{x}) = \sum_{k=1}^K \beta_k\Phi_k(x_k, \textbf{x}_{-k}).
\end{equation}
A global optimal point of $\Phi$ is a Pareto point {\cite{Miettinen}}.

Now we take $\Phi_k = \min(1,\alpha_k)$, where $\alpha_k$ has been defined as before. The reason for choosing $\min(1,\alpha_k)$ instead of $\alpha_k$ is that taking $\alpha_k$ will allow the possibility of the Pareto point to maximize the sum of the $\alpha$'s while some $\alpha_k$ may be strictly less than $1$, even when it was possible to satisfy the QoS of all the users. This will not happen with the modified utility function.

Since the strategy space $\mathscr{X} = \prod_{k=1}^K \mathscr{X}_k$ is compact and each $\Phi_k$ for our problem is continuous, $\Phi$ has a global maximum and hence a Pareto point. The same can be said about the cost function ($C_k = -\Phi_k$).

As in case of CCE, if we have enough resources to satisfy the QoS of all the users in each FC, then we will get $\alpha_k = 1$ for all the FCs. Now the rates received by some of the users may be more than needed and hence the powers used by an FC may be more than needed. Thus, now as per the CCE case we keep the channel allocations fixed and play the power game to reduce the powers needed by each FC. Being a potential game, it also provides a Pareto point.

Instead of using (10), we can also use (12) to get Pareto points. Now, as said for CCE, we will not need to play a power game later on. Thus, the complexity to obtain the Pareto point is less. However, we expect that Pareto point obtained via (10) fairer solution within a FC (higher $\alpha_k$). We will see this in example provided in the section VII.

To obtain a global optimal of (14) in our distributed setup we use the following algorithm. We descretize the power levels. Then the overall strategy space becomes a finite set. We use Algorithm 3 below to get its global optimal. If each $\Phi_k$ is continuous and the strategy space is compact, then we can easily show that for each $\epsilon > 0$ there is a discretization of state space for which the global optimal is within $\epsilon$ of the global optimal of $\Phi$.

In Algorithm 3, each user updates its strategy as follows. It picks randomly, uniformly one of its channels and changes its allocation randomly, to one of the other users. Power allocation to all the channels are randomly, uniformly picked from the discrete set. If at these new strategies of all the FCs, $\Phi(\textbf{x})$ increases, then the users update to the new strategy; otherwise they retain the previous one.

We can modify Algorithm 3 by changing the first step to change the allocation of all the channels (instead of only one of the channels), uniformly to the users. We call this Algorithm 3M.

Proposition 1 below shows that Algorithm 3M converges to a Pareto point with probability 1. However, Algorithm 3, which is a heuristic (e.g., does not guarantee convergence to a Pareto point) often obtains the Pareto point much faster than Algorithm 3M. We will show this in section VII via examples. The reason for this is that Algorithm 3 does local stochastic search instead of global stochastic search. Global search ensures convergence but since it searches over a much larger space, it can be very slow in finding better strategies. However, the local search can get struck into local optima. But in our setup, this local search is not so local (due to multiple users and FCs) and hence chances of our obtaining a global optimum are high. Indeed in most of the cases we considered we obtained the global optimum.
\begin{theorem1}
The number of steps needed by Algorithm 3M to reach a Pareto optimal point $\textbf{x}^*$ with probability $\ge 1-\epsilon$ is
\begin{equation*}
n \leq \frac{\log \epsilon}{\log(1 - \frac{1}{M})},
\end{equation*}
where $\epsilon \in (0,1]$ and $ M = |\mathscr{X}|$, the cardinality of $\mathscr{X}$. Also, $P[R<\infty]=1$, where $R$ is the number of steps to reach $\textbf{x}^*$.
\end{theorem1}
\begin{proof}
Let $\textbf{x} \in \mathscr{X}$ be picked with uniform distribution. Then,
\begin{equation*}
P(R \leq n) = \sum_{i=1}^n \frac{1}{M}(1 - \frac{1}{M})^{i-1}
\end{equation*}
\begin{equation}
           = 1 - (1 - \frac{1}{M})^n.
\end{equation}
Hence, $P[R<\infty]=\lim_{n\rightarrow\infty}P[R\leq n] = 1$. Also, for any given $\epsilon \in (0,1]$, probability of hitting $\textbf{x}^*$ in $n$ steps is $\geq 1- \epsilon$, if
\begin{equation*}
1 - (1 - \frac{1}{M})^n \geq 1-\epsilon
\end{equation*}
which implies, $n \leq \frac{\log \epsilon}{\log(1 - \frac{1}{M})}.$
\end{proof}
From (15), $R$ has a finite exponential moment.

In this algorithm (3 or 3M), the random updates of all the users are taken into account together and $\Phi$ is computed by each user and the new strategy is decided. Thus, each user needs to know the new updated strategy of all users, as for the NE before but also their utility functions. Alternately, all users send their new strategy to one of the users or a central observer who computes $\Phi$ and announces to all whether to keep the new update or the previous. The computational complexity of this algorithm is much less than that of Algorithm 2 because of the expected value computation in Algorithm 2.

We can also add a termination condition. If the strategy $\textbf{x}^n$ stays same for $N_1>0$ consecutive iterations, we stop. The probability that we stop at the global optimum increases with $N_1$.

The Pareto point obtained here may have some fairness issues, i.e., although the point obtained maximizes $\Phi$, some player may get larger $\alpha_k$ than others. This can be adjusted by picking Pareto points which are also Nash bargaining solutions {\cite{roger}}. Such solutions are discussed below.
\vspace{1cm}
\begin{algorithmic}
\begin{algorithm}
\caption{Algo-Pareto 1}
\STATE Initialization: All the FCs $k=1, 2, ..., K$ will pick an arbitrary strategy denoted by $\textbf{x}^0 \in \mathscr{X}$ and exchange with each other.
\STATE 1. Each FC $k$ picks one of the channels randomly and reallocates to one of the users uniformly at random. Then it randomly reallocates its powers to each channel. Let the new overall strategy be denoted by $\textbf{x}$. Then,
\[\textbf{x}^n = \left \{
\begin{array} {ccl}
\textbf{x}, & &\textrm{if ($\Phi(\textbf{x}) > \Phi(\textbf{x}^{n-1})$}),\\
         \textbf{x}^{n-1}, & & otherwise.
\end{array}
 \right. \]
\STATE 2. Repeat until converges or the termination criterion is met.
\end{algorithm}
\end{algorithmic}
\section{Nash Bargaining Solution}
In addition to not being fair, usually there are multiple Pareto points. As opposed to this, a Nash Bargaining (\cite{roger}) solution guarantees a fair, Pareto optimal solution which is unique. The following theorem provides existence of a Nash Bargaining solution.
\begin{theorem}({\cite{roger}})
Let the set of possible payoffs $\textsl{F}\subseteq \textrm{R}^K$ be a closed, convex set. Let $\boldsymbol{\hat{\alpha}}=(\hat{\alpha}_1, \hat{\alpha}_2, ..., \hat{\alpha}_K)$ be the payoffs that the players would expect if they did not cooperate. Assume that $\{\boldsymbol{\alpha} \in \textsl{F}: \alpha_k\geq \hat{\alpha}_k, \forall k \in \mathscr{I}\}$ is a nonempty bounded set. Also assume that there exists some $\boldsymbol{\alpha} \in \textsl{F}$ such that $\alpha_k>\hat{\alpha}_k, \forall k$. Then there exists a unique $\boldsymbol{\alpha} \geq \boldsymbol{\hat{\alpha}}$ that maximizes
\begin{equation}
%\Phi(\textsl{F}, \textbf{v}) \in \arg\max_{\textbf{x}\in \textsl{F}, x_k\geq v_k, \forall k} \prod_{k\in \mathscr{I}}(x_k - v_k)
\prod_{k\in \mathscr{I}}(\alpha_k - \hat{\alpha}_k),
\end{equation}
which is the Nash Bargaining Solution of the problem. $\hspace{6cm}\Box$
\end{theorem}

We consider the problem in (10)-(11) for computation of Nash Bargaining solution. A feasible pay off allocation is $\textsl{F} = \{\boldsymbol{\alpha}=(\alpha_1, \alpha_2, ..., \alpha_K): 0 \leq \alpha_k \leq 1, \forall k\}$ which is a closed, convex set. Also consider $\boldsymbol{\hat{\alpha}}$ such that there is a feasible $\boldsymbol{\alpha}>\boldsymbol{\hat{\alpha}}$. Then from the above Theorem 1 we have a unique Nash Bargaining solution $\Phi(\textsl{F},\boldsymbol{\hat{\alpha}})$. Equation (16) can be rewritten as (by taking logarithm)
\begin{equation*}
%\Phi(\textsl{F}, \textbf{v}) \in \arg\max_{\textbf{x}\in \textsl{F}, x_k\geq v_k, \forall k} \prod_{k\in \mathscr{I}}(x_k - v_k)
\Phi(\textsl{F}, \boldsymbol{\hat{\alpha}}) \in \operatorname*{arg\,max}_{\boldsymbol{\alpha}\in \textsl{F}, \alpha_k\geq \hat{\alpha}_k, \forall k} \sum_{k\in \mathscr{I}}\log(\alpha_k - \hat{\alpha}_k).
\end{equation*}
In order to avoid a non-positive number within logarithm,we consider
\begin{equation}
%\Phi(\textsl{F}, \textbf{v}) \in \arg\max_{\textbf{x}\in \textsl{F}, x_k\geq v_k, \forall k} \prod_{k\in \mathscr{I}}(x_k - v_k)
\Phi(\textsl{F}, \boldsymbol{\hat{\alpha}}) \in \operatorname*{arg\,max}_{\boldsymbol{\alpha}\in \textsl{F}, \alpha_k\geq \hat{\alpha}_k, \forall k} \sum_{k\in \mathscr{I}}\log(c+\alpha_k - \hat{\alpha}_k)
\end{equation}
where $c>0$ is appropriately chosen. We use Algorithm 3/3M to compute the Nash Bargaining Solution via (17).

$\boldsymbol{\hat{\alpha}}$ should be carefully chosen. We can take $\boldsymbol{\hat{\alpha}}=0$. We can also choose it as a NE or according to some QoS requirement.
\section{Coexistence of Voice and Data Users}

In this section we consider the case where each FC $k$ may have data users and voice users. We extend our algorithms developed above to this scenario and compute a CCE, a Pareto optimal point and a Nash Bargaining solution.

Let $D^k$ be the set of data users and $V^k$ be the set of voice users. We consider the following modified problem. At each FC $k$,
\begin{equation}
\max \quad \alpha_k
\end{equation}\vspace{-7mm}\\
such that \vspace{-3mm}\\
\begin{equation}
\sum_{i=1}^N{C}^k_{i,j} A_{i,j}^k \geq \alpha_k \bar{R}^k_j, \quad \forall j \in D^k,
\end{equation}
\begin{equation}
\sum_{i=1}^N{C}^k_{i,j} A_{i,j}^k \geq \bar{R}^k_j, \quad \forall j \in V^k,
\end{equation}
and
\vspace{-3mm}
\begin{equation}
\sum_{j=1}^{M_k}A_{i,j}^k \leq 1, \quad \forall i=1, 2, ..., N.
\end{equation}
Equation (19) specifies that a largest fraction of the QoS of data users are satisfied and equation (20) specifies that QoS of voice users are fully satisfied. Since voice users require much less rate, the network will often have enough resources to satisfy (19).

To obtain a CCE for this problem we use the algorithms of Section III only but change the utilities $\alpha_k$ of FC $k$ as follows: if at a given power and channel allocation, the rate of any of its voice users is not satisfied, then we make its $\alpha_k=0$; otherwise compute it as in Section III based on data users only. This will ensure that if an FC has a positive utility, the rate requirements of all its voice users is satisfied.

Next we consider the Pareto points. Using the modified $\alpha_k$ given above does not always provide a solution with stochastic local search. It can be made to work if we carefully tailor our stochastic local search to the constraints. Thus, instead,  we consider penalty function method corresponding to eq. (12). Combining equations (19) and (20), as in equation (12) we consider
\begin{equation*}
\frac{1}{|V^k|}\sum_{j \in V^k}\min\{\frac{1}{\bar{R}^k_j}\sum_{i=1}^N{C}^k_{i,j} A_{i,j}^k,1 \} + \min_{j \in D^k}\{\frac{1}{\bar{R}^k_j}\sum_{i=1}^N{C}^k_{i,j} A_{i,j}^k \}
\end{equation*}
\begin{equation}
\geq 1+\alpha_k.
\end{equation}
and Algorithm 2A to obtain a CCE.

Next we consider LHS of equation (22) as $\Phi_k$ and obtain a Pareto optimal solution from equation (14) and Algorithm 3 and 3M. Also, $\Phi_k$ is closed and convex. Hence it has a Nash Bargaining solution. By considering $\alpha_k = \Phi_k$ in equation (17), we obtain a Nash Bargaining Solution. For all voice and data users (whose QoS is satisfied) at their allocated channels we can minimize power using KKT conditions while ensuring that their rate requirements are satisfied.

From equation (12), we formulate equations (18)-(21) as,
\begin{equation*}
C_k(x_k,\textbf{x}_{-k}) = \sum_{i=1}^N P^k_i + \rho \sum_{j \in D^k}\max\{0,\bar{R_j}^k-\sum_{i=1}^NC^k_{i,j} A^k_{i,j}\}
\end{equation*}
\begin{equation}
+ \rho \sum_{j \in V^k}\max\{0,\bar{R_j}^k-\sum_{i=1}^NC^k_{i,j} A^k_{i,j}\}
\end{equation}
Using equation (23) and (14), we obtain Pareto optimal points.
%**********************************
\section{Examples}
In this section we consider a system with $2$ FCs each with $2$ users and $4$ subchannels deployed in a MC. We consider Time Division Duplex (TDD) scenario where all parameters, such as rate requirements, interference matrix and subchannel gain matrices do not change significantly within a given TDD duration. We use the following parameters:\vspace{1mm}\\
Subchannel bandwidth B = $180kHz$, SNR Gap $\Gamma$ = $1$, Noise Power $N_0 = 1*10^{-9}$W, Noise variance $\sigma^2 = N_0B = 1.8*10^{-4}$watt-sec, $\rho=15$, $\beta_k=1$ for $k=1, 2$.\\
Subchannel gain matrices; $G^1, G^2, G^{1,2}, G^{2,1}$ were obtained by sampling from Gaussian distributions $\mathscr{N}(1,1), \mathscr{N}(0.5,1), \mathscr{N}(0,1), \mathscr{N}(0,1)$ and taking square. The interference matrices (mw) are \\
\[I^1 = \left(
                                  \begin{array}{cc}
                                      4  &  3  \\
     11  &  4  \\
     8  &   9  \\
    9  &   13  \\
       \end{array}
                                \right), \hspace{1mm}I^2 = \left(
                                  \begin{array}{cc}
                                     10  &   1  \\
    11  &   6  \\
     6  &   1  \\
     1  &  14  \\
              \end{array}
       \right)\]\\
Maximum powers (mw) allocated to the subchannels are,\\
$\bar{P}^1 = [2.1, 1.2, 8.0, 1.8], \bar{P}^2 = [1.1, 1.5, 7.5, 1.9]$.\\
%\[\bar{P}^1 = \left[\begin{array}{cccc}
%           2.1, & 1.2, & 8.0, & 1.8
%           \end{array}\right],\]
%\[\bar{P}^2 = \left[\begin{array}{cccc}
%           1.1, & 1.5, & 7.5, & 1.9
%           \end{array}\right].\]
Rate requirements of the users, in kbps, in the two FC are,\\
\hspace{4cm}$\bar{R}^1 = [200, 220], \bar{R}^2 = [160, 180]$.
 %\[\bar{R}^1 = \left[\begin{array}{cc}
%                                                                      200, & 220\end{array}\right],\]
%                                       \[\bar{R}^2 = \left[\begin{array}{cc}
%                                        160, & 180 \end{array}\right].\]
%\begin{eqnarray*}
%\bar{R}^1=[200, 220], \bar{R}^1=[160, 180].
%\end{eqnarray*}
\begin{figure}
  \centering
  \includegraphics[scale=0.2]{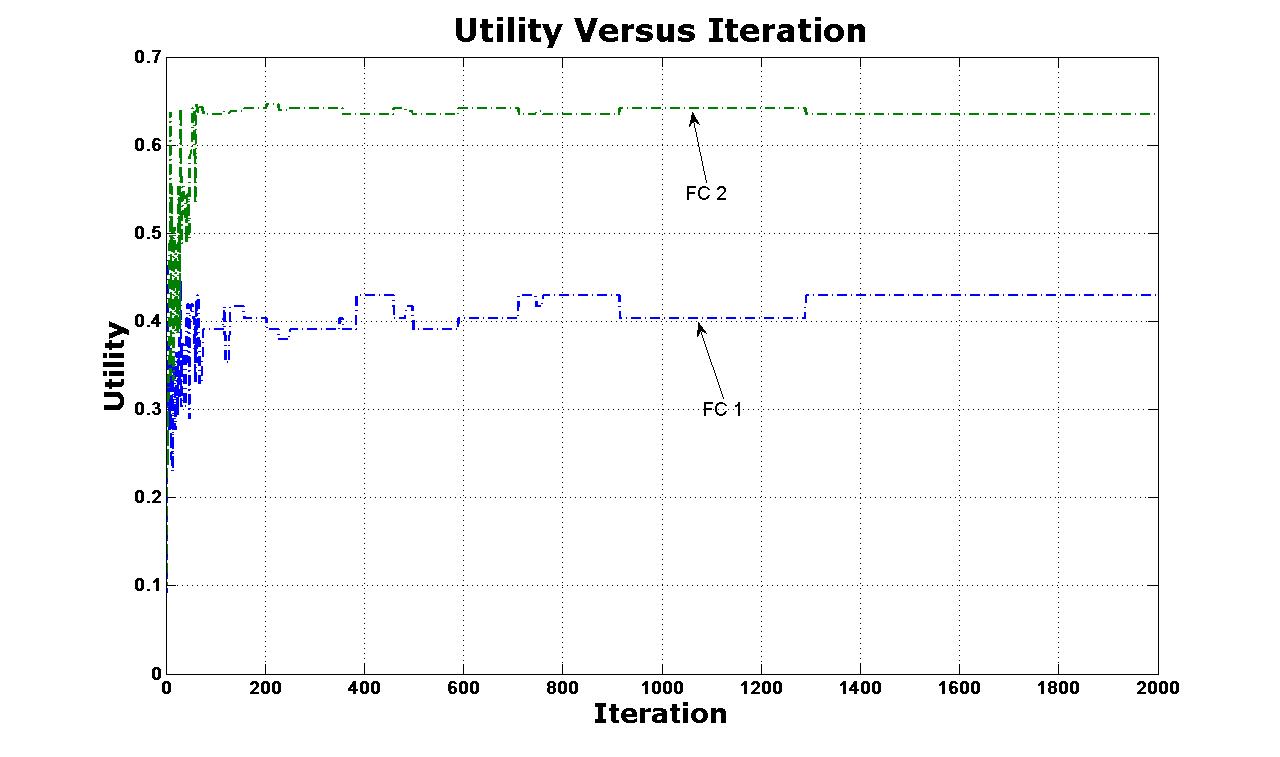}\\
  \vspace{-4mm}
  \caption{Example 1, Correlated Equilibrium using Algorithm-3 of \cite{uday1}.}
  %\vspace{-6mm}
\end{figure}
\vspace{-0.7mm}

We have verified that the rate requirement of all the users cannot be satisfied for this case. Thus, we obtain a fair NE
via Algorithm 3 (regret matching) of \cite{uday1}. There are multiple NE. Depending on the initial conditions, the algorithm may converge to a different NE. We
show the convergence of this algorithm in Fig 1. We have observed that the regret algorithm converges to the same NE in this example even though we started with different initial conditions. For this NE, the subchannel
allocation for the two FC's converges to $CA^1 = [2, 1, 1, 1], CA^2 = [2, 2, 1, 2]$ and the power allocations are $P^1$(in mw)$ = [2.1, 1.2, 8.0, 1.8], P^2$(in mw)$ = [1.1, 1.5, 7.5, 1.9]$  with the
utilities $\alpha_k$ as $0.4294, 0.6348$ and the allocated rates (in kbps) $[85.8, 97.9], [123.5, 114.2]$.

Using the multiplicative weights Algorithm (Algorithm 1), we obtain the same CCE. The convergence of utilities is shown in Fig 2.

\begin{figure}
  \centering
  \includegraphics[scale=0.2]{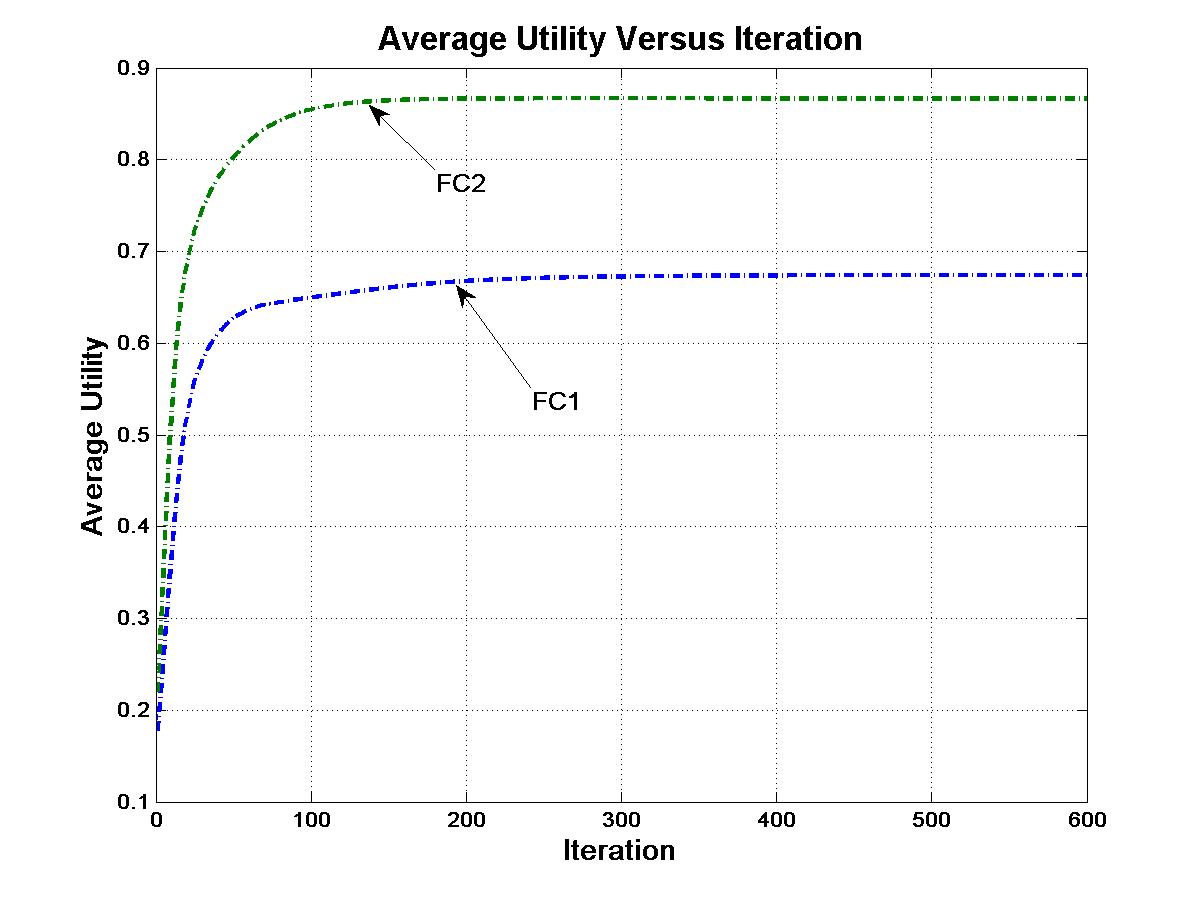}\\
  \vspace{-4mm}
  \caption{Example 1, CCE using Algorithm-1 for problem (10).}
  %\vspace{-6mm}
\end{figure}

For cost (13), using Algorithm 2A, we get a CCE with the subchannel allocations $CA^1 = [2, 2, 1, 1], CA^2 = [2, 2, 1, 2]$ and the power allocations $P^1$(in mw)$ = [2.1, 1.2, 8.0, 1.8], \\P^2$(in mw)$ = [1.1, 1.5, 7.5, 1.9]$  with allocated rates (in kbps) $[78.1, 106.4], [123.5, 114.2]$. Convergence of the algorithm is shown in Fig 3. We see that compared to the CCE obtained via (10), this CCE is less fair (in terms of fraction of rates obtained) to the users in FC 1.

It is observed from Figs 1, 2 and 3 that time taken to converge to a CE by using Algorithm 3 of \cite{uday1} is larger (around 1300 iterations) as compared to the CCE computation using Algorithm 1 (around 200 iteration for problem 10 and around 400 iteration for problem 13, Algorithm 2A). Moreover, computations per iteration for Algorithms 1 and 2A are much less than for Algorithm 3 of \cite{uday1}.
\begin{figure}
  \centering
  \includegraphics[scale=0.2]{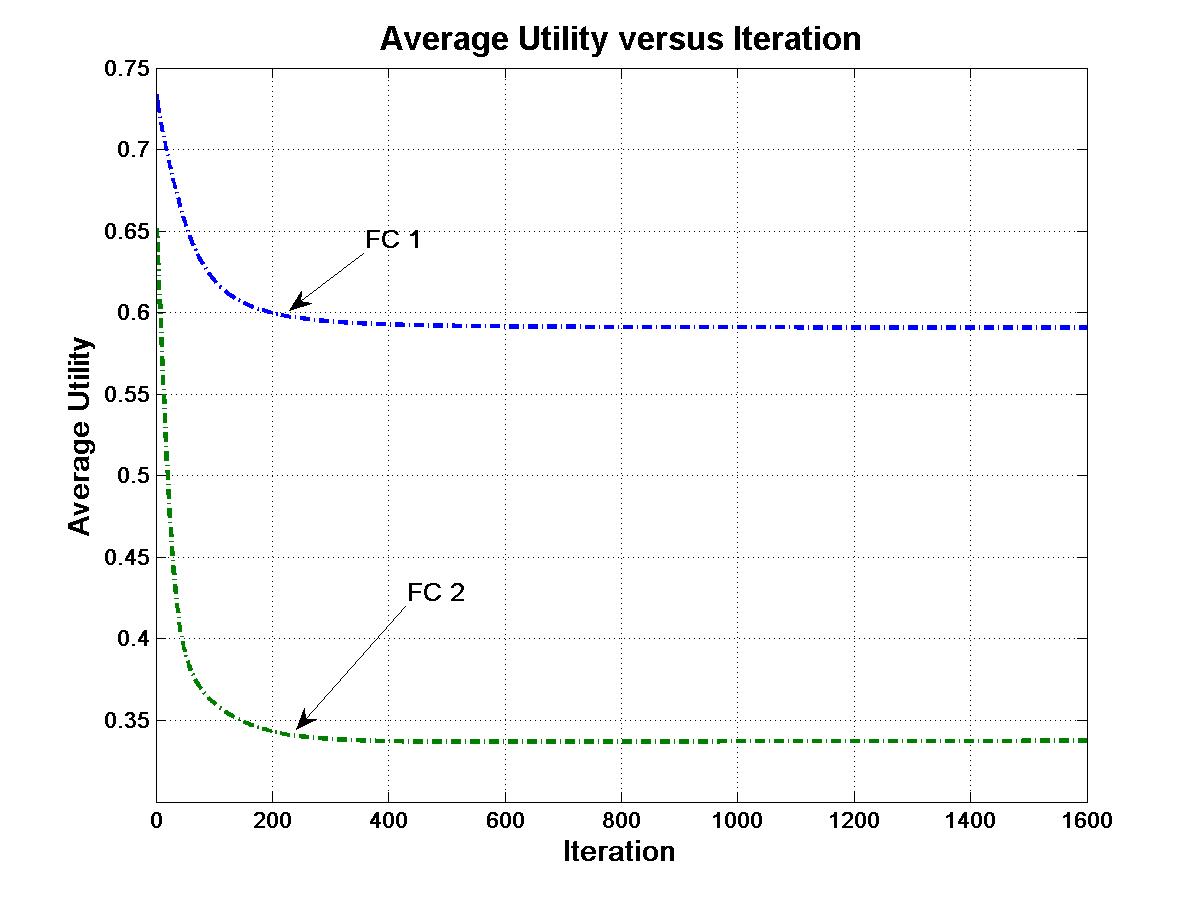}\\
  \vspace{-4mm}
  \caption{Example 1, CCE using Algorithm-2A for problem (13).}
  %\vspace{-6mm}
\end{figure}

Convergence to a Pareto optimal point for utilities $\alpha_k$, obtained using Algorithm 3 for (10)-(11) is shown in Fig 4. At this point the subchannel allocations are $CA^1 = [2, 2, 1, 1], \\CA^2 = [2, 2, 1, 2]$ and the power allocations are $P^1$(in mw)$ = [1.4, 1.2, 8.0, 1.8], P^2$(in mw)$ = [1.1, 1.5, 7.5, 1.9]$  with allocated rates (in kbps) $[78.1, 77.7], [123.5, 125.9]$ with utilities $[0.3532, 0.6994]$.
\begin{figure}
  \centering
  \includegraphics[scale=0.2]{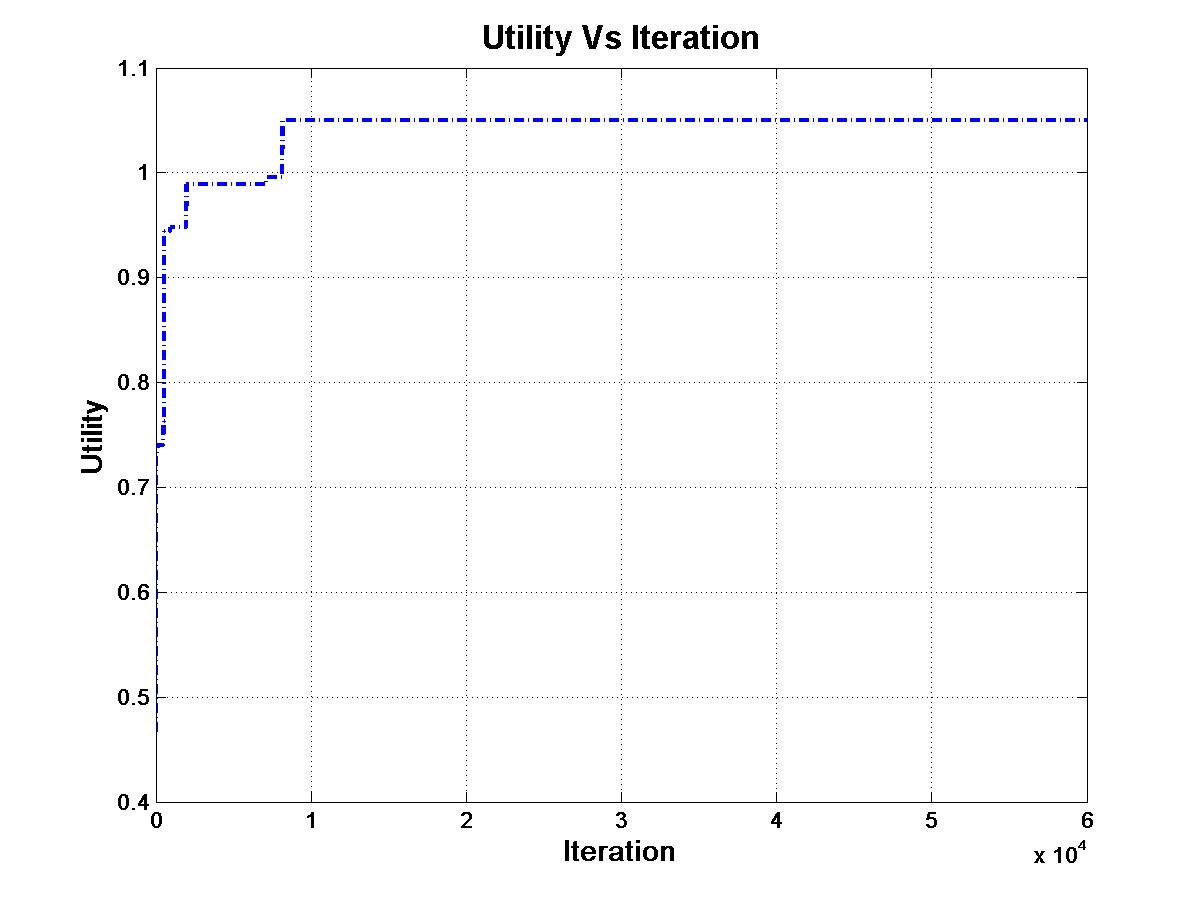}\\
  \vspace{-4mm}
  \caption{Example 1, Pareto Optimal using Algorithm-3 for problem (10).}
  %\vspace{-6mm}
\end{figure}

Convergence to a Pareto optimal point for utilities $\alpha_k$, obtained using Algorithm 3M for (10)-(11) is shown in Fig 5. At this point the subchannel allocations are $CA^1 = [2, 2, 1, 1], CA^2 = [2, 2, 1, 2]$ and the power allocations are $P^1$(in mw)$ = [1.4, 1.2, 8.0, 1.2], P^2$(in mw)$ = [1.1, 1.5, 7.5, 1.9]$  with allocated rates (in kbps) $[70.7, 77.7], [123.5, 126.0]$ with utilities $[0.3532, 0.70]$. It is observed that this algorithm takes around $5.5\times 10^5$ iterations to converge to a Pareto optimal while Algorithm 3 took around $0.9\times 10^4$ iterations. We ran both the algorithms several times and found that Algorithm 3M almost always took much longer to converge.
\begin{figure}
  \centering
  \includegraphics[scale=0.2]{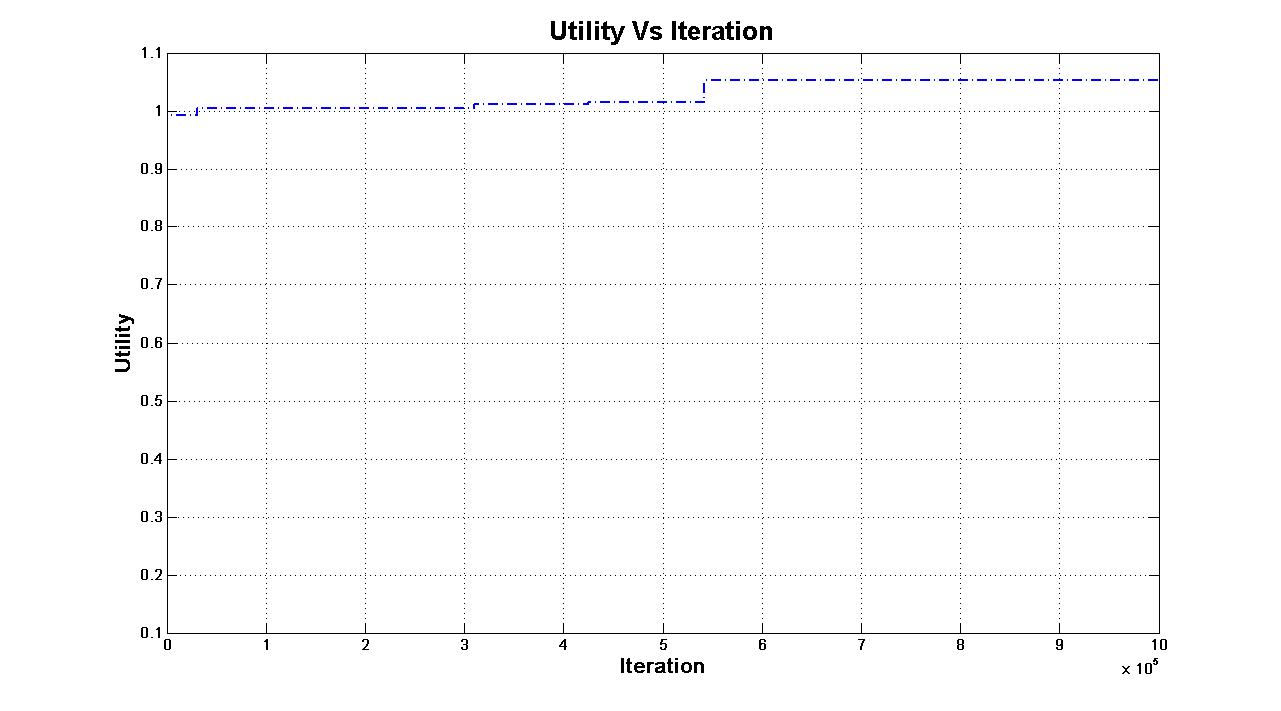}\\
  \vspace{-4mm}
  \caption{Example 1, Pareto Optimal using Algorithm-3M for problem (12)with uniform channel selection.}
  %\vspace{-6mm}
\end{figure}

Convergence to a Pareto optimal point obtained using Algorithm 3 for (12) with $\rho=5$ is shown in Fig 6. We get a Pareto optimal point with subchannel allocations $CA^1 = [2, 1, 1, 2], CA^2 = [2, 2, 1, 1]$ and the power allocations are $P^1$(in mw)$=[2.1, 0.8, 8.0, 0.0], P^2$(in mw)$=[1.1, 1.5, 5.0, 1.9]$  with allocated rates (in kbps) $[74.8, 97.9],$ $[177.6, 87.8]$. The corresponding $\alpha_k$ obtained is $[0.3740, 0.4878]$. This Pareto point is less fair in FC2, compared to the Pareto point obtained via (10).

\begin{figure}
  \centering
  \includegraphics[scale=0.2]{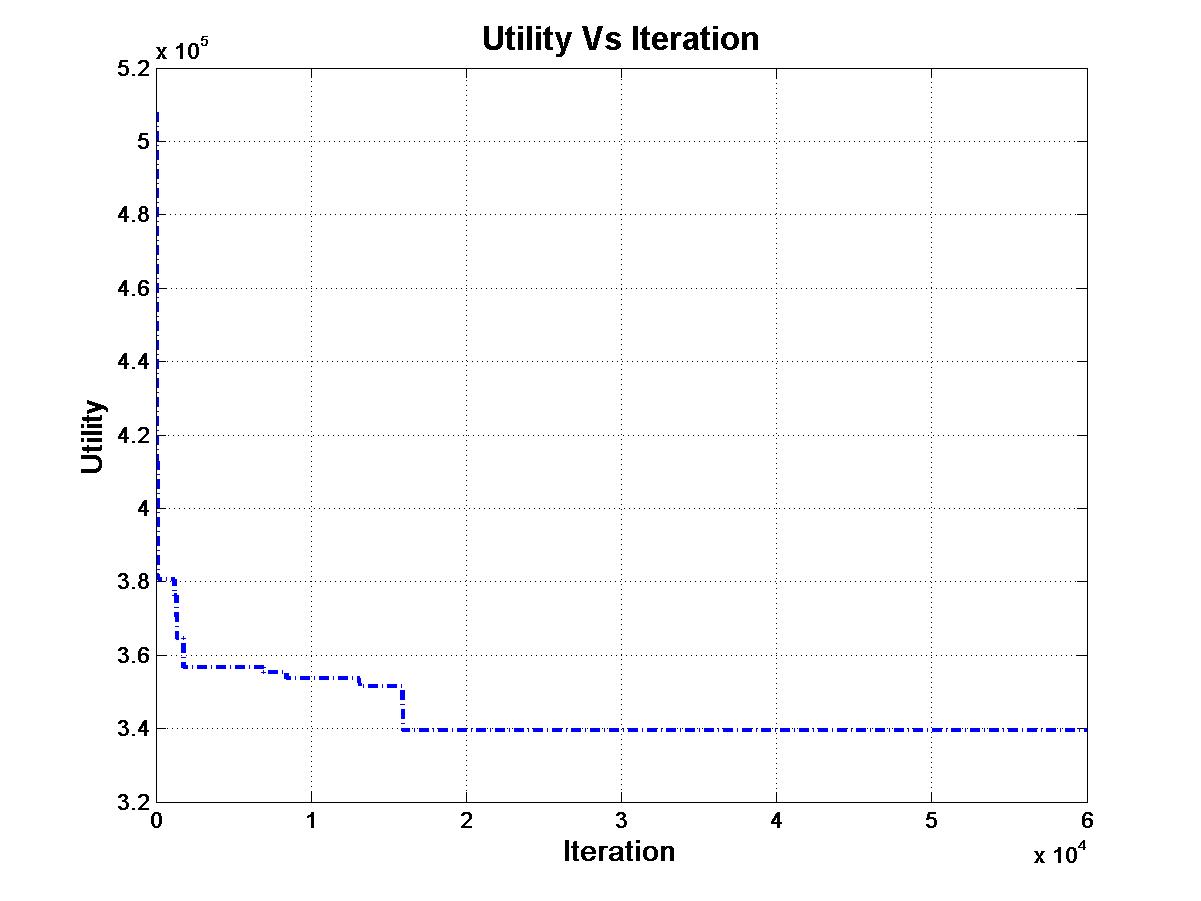}\\
  \vspace{-4mm}
  \caption{Example 1, Pareto point using Algorithm-3 for problem (12).}
  %\vspace{-6mm}
\end{figure}
Utilities obtained via Nash Bargaining solutions with $\boldsymbol{\hat{\alpha}}=\{0.42, 0.42\}$ using (17) are shown in Fig 7. The subchannel allocation is $CA^1 = [2, 2, 1, 1], CA^2 = [2, 2, 1, 2]$ and the power allocations are $P^1$(in mw)$ = [2.1, 0.8, 8.0, 1.8], P^2$(in mw)$ = [1.1, 1.0, 5.0, 1.9]$  with allocated rates (in kbps) $[92.3, 103.8], [88.5, 106.2]$ with utilities $[0.3532, 0.6994]$. Here, we observe that power allocated in FC 2 is reduced.

It is observed that for (10-11), $\alpha_1+\alpha_2$ at the Pareto point is $1.0526$, which is larger than that of the Nash Bargaining solution obtained, with 1.0146. Also, $|\alpha_1-\alpha_2|$ at NB solution is 0.0916 but for the Pareto point is 0.3423. This indicates that NB solution is a fairer solution than a Pareto point. The Pareto point for (12) provides $\alpha_1+\alpha_2 = 0.8618$ which is less compared to the Pareto point obtained via (10-11).
\begin{figure}
  \centering
  \includegraphics[scale=0.2]{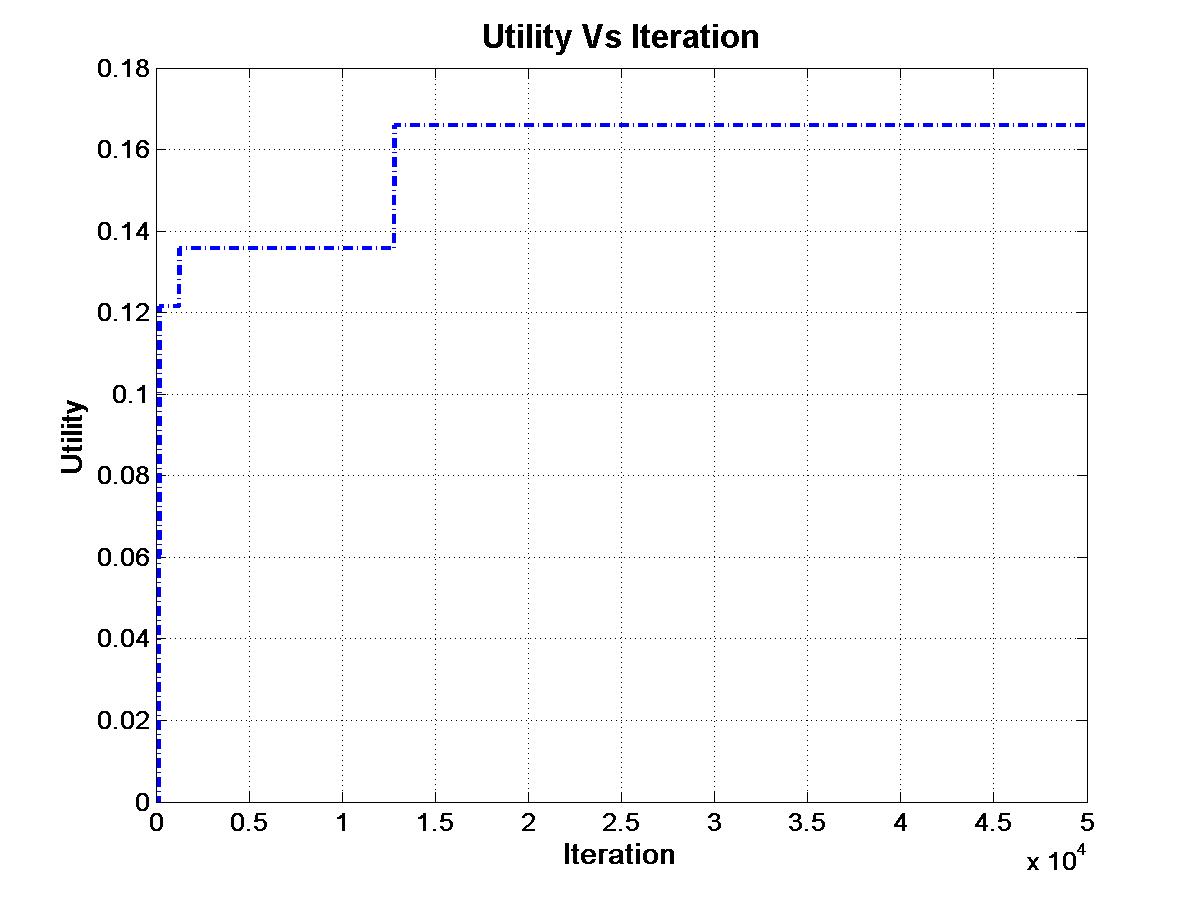}\\
  \vspace{-4mm}
  \caption{Example 1, Nash Bargaining solution using Algorithm-3 for problem (17).}
  %\vspace{-6mm}
\end{figure}

Next we consider an example when the QoS of all the users in both the FCs can be satisfied. We generated the subchannel gain matrices $G^1, G^2, G^{1,2}, G^{2,1}$ as before.
The interference matrices (mw) are \\\[I^1 = \left(
                                  \begin{array}{cc}
                                      11  &  13  \\
     3  &  14  \\
     7  &   8  \\
    10  &   2  \\
       \end{array}
                                \right),\hspace{1mm} I^2 = \left(
                                  \begin{array}{cc}
                                     2  &   10 \\
    7  &   11  \\
     6  &   4  \\
     9  &  10  \\
              \end{array}
       \right)\]\\
Maximum powers (mw) allocated to the subchannels are,
%$\bar{P}^1 = [ 15, 15, 15, 15], \bar{P}^2 = [ 15, 15, 15, 15]$.\\
\[\bar{P}^1 = \left[\begin{array}{cccc}
           15, & 15, & 15, & 15
           \end{array}\right], \bar{P}^2 = \left[\begin{array}{cccc}
           15, & 15, & 15, & 15
           \end{array}\right].\]
Rate requirement of the users, in kbps, are,
%$\bar{R}^1 = [225, 100], \bar{R}^2 = [200, 150]$.\\
 \[\bar{R}^1 = \left[\begin{array}{cc}
                                                                      225, & 100\end{array}\right], \bar{R}^2 = \left[\begin{array}{cc}
                                        200, & 150 \end{array}\right].\]
For the problem (10)-(11) using Algorithm 1, we obtain a CCE with the following specifications. The convergence of utilities is shown in Fig 8. The subchannel allocation is $CA^1 = [1, 1, 2, 1], CA^2 = [2, 2, 1, 2]$ and the power allocations are $P^1$(in mw)$ = [15, 0, 8.8, 13.2], P^2$(in mw)$ = [0, 0, 9.2, 11.8]$. Now $\alpha_k=1$ for both the FCs and the allocated rates are same as the required rates. Total allocated powers (in mw) in each FC are $37.0, 21.0$ respectively.
\begin{figure}
  \centering
  \includegraphics[scale=0.2]{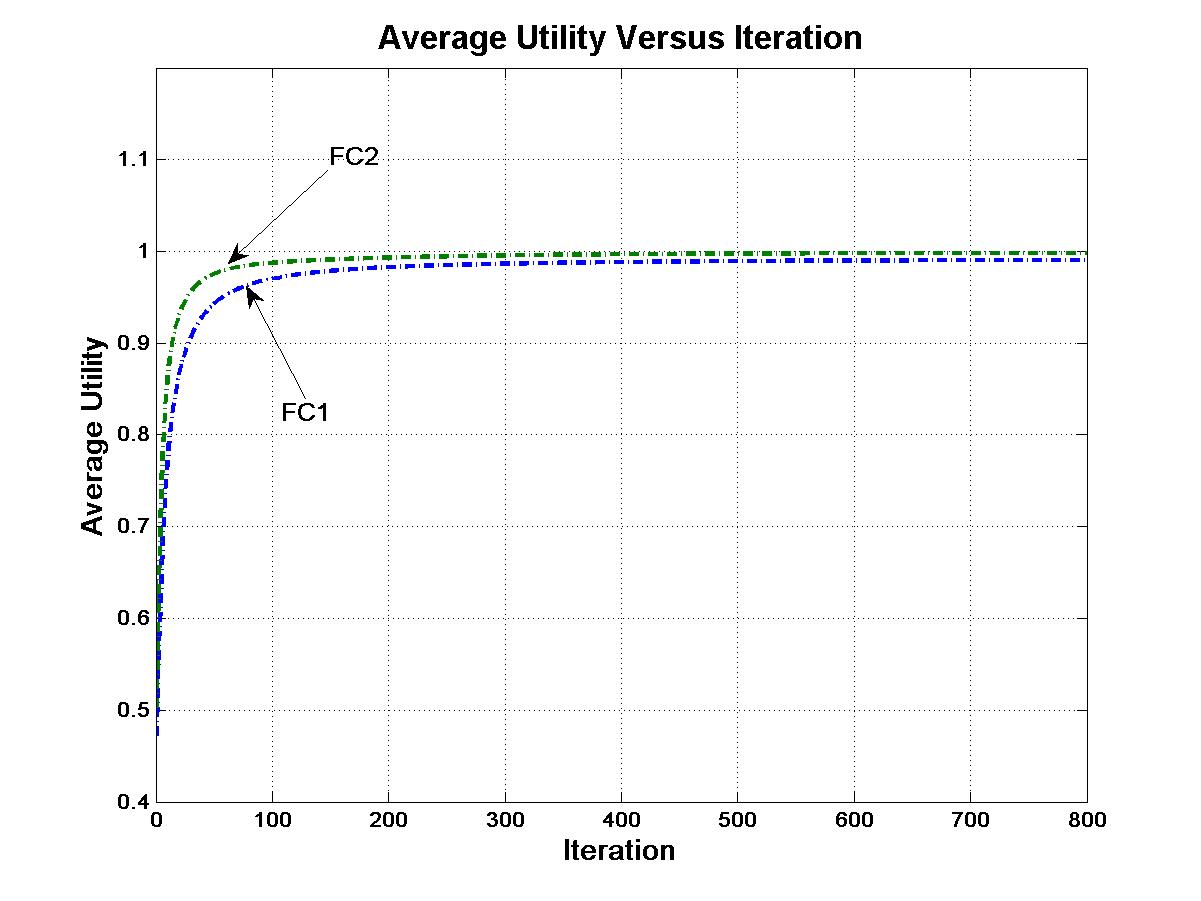}\\
  \vspace{-4mm}
  \caption{Example 2, CCE using Algorithm-1 for eq (10).}
  %\vspace{-6mm}
\end{figure}

Similarly, for the problem (13), we obtained a CCE with subchannel allocations $CA^1 = [1, 1, 2, 1], CA^2 = [2, 2, 1, 2]$, power allocations $P^1$(in mw)$ = [15, 5, 10, 10], P^2$(in mw)$ = [5, 0, 10, 10]$  with allocated rates (in kbps) $[273.2, 111.0], [207.7, 151.6]$. Total allocated powers (in mw) in the two FCs are $40.0, 25.0$. The convergence of the utilities is shown in Fig 9.

It is observed that CCEs obtained via (10) and (13) give the same channel allocation, whereas power allocation is slightly different. This is because for (10) we do power control using the obtained channel allocation.
\begin{figure}
  \centering
  \includegraphics[scale=0.2]{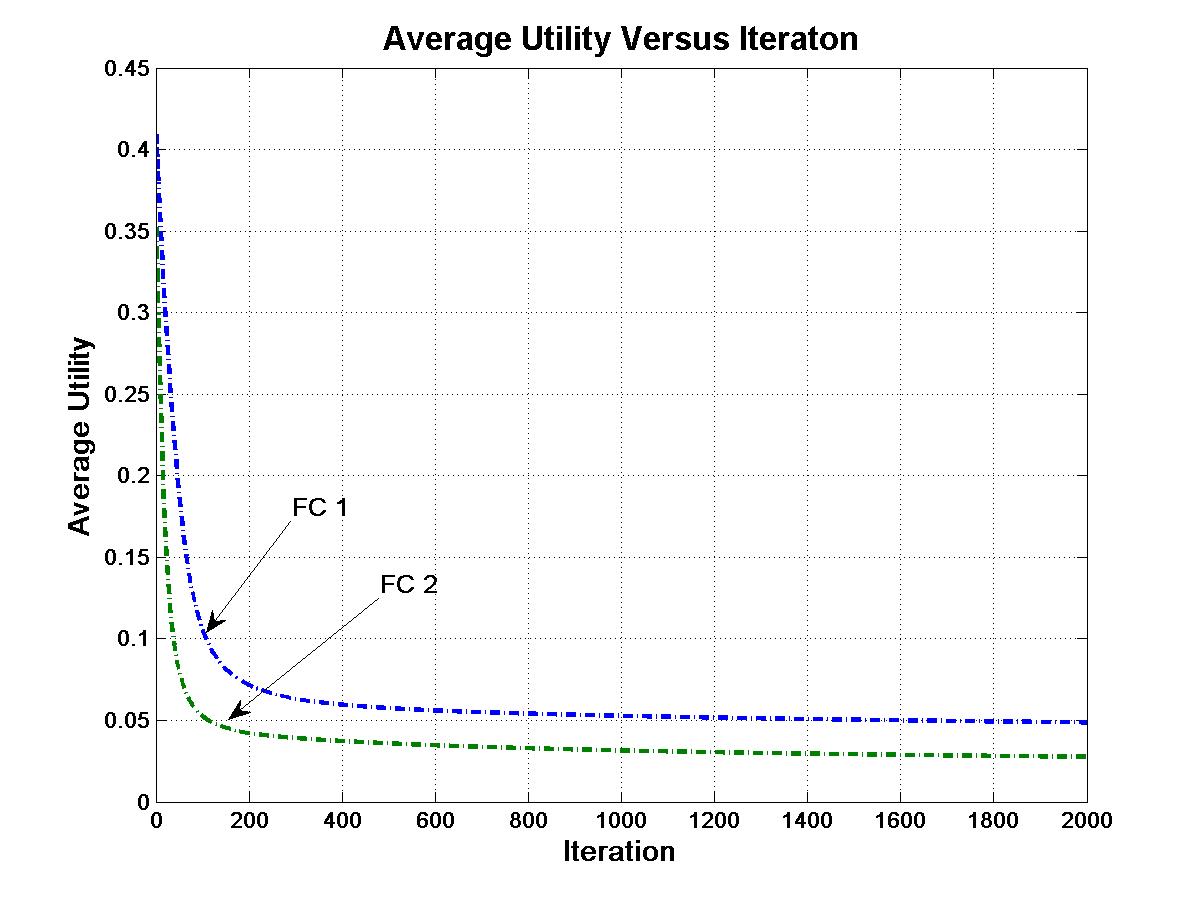}\\
  \vspace{-4mm}
  \caption{Example 2, CCE using Algorithm-1 for eq (13).}
  %\vspace{-6mm}
\end{figure}

For the problem (10)-(11) using Algorithm 3, we obtain the Pareto point with the convergence plot shown in Fig 10. The Pareto point has subchannel allocations $CA^1 = [1, 1, 2, 1], CA^2 = [1, 2, 2, 1]$ and the power allocations $P^1 = [9.6, 13.3, 9.9, 0.0], P^2 = [0.0, 0.0, 9.5, 10.0]$  with the allocated rates same as the required rates. Total allocated powers (in mw) are $32.9, 19.5$. We see that these powers are less than for the two CCEs obtained above.
\begin{figure}
  \centering
  \includegraphics[scale=0.2]{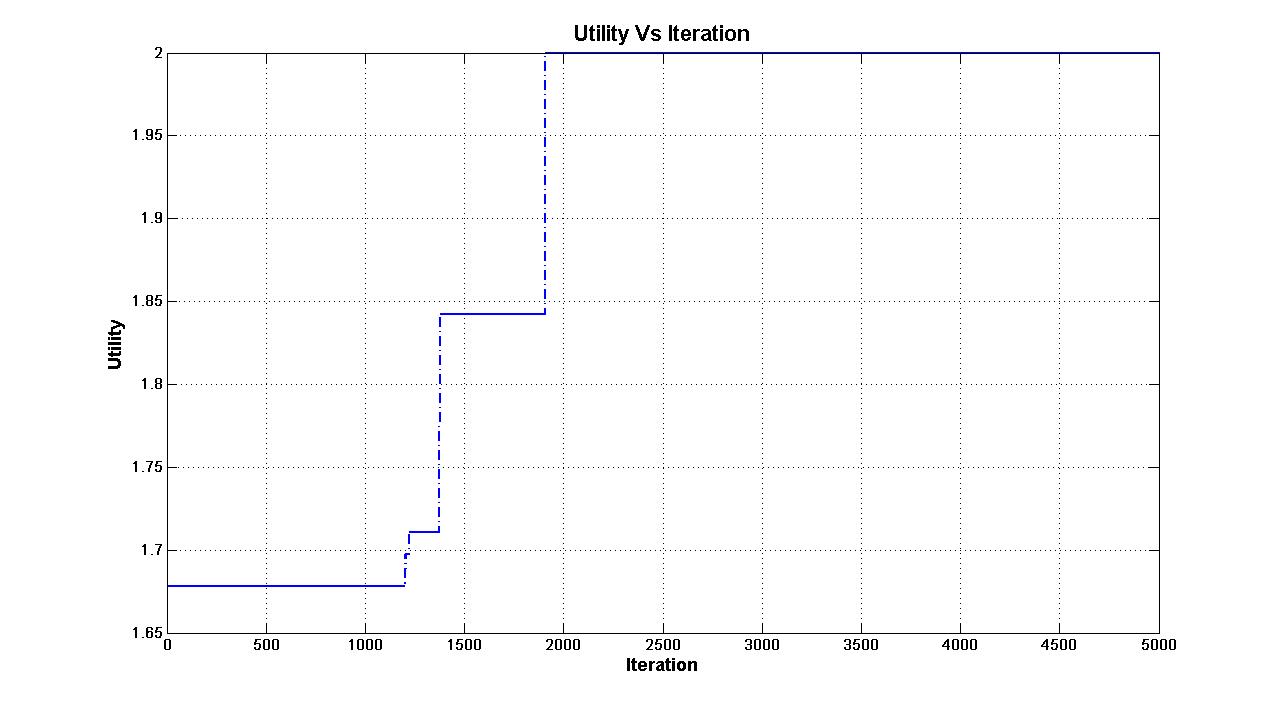}\\
  \vspace{-4mm}
  \caption{Example 2, Pareto point using Algorithm-3, eq (10).}
  %\vspace{-6mm}
\end{figure}

Utilities of the Pareto points obtained via (12) with $\rho=5$ are shown in Fig 11. The subchannel allocation is $CA^1 = [1, 1, 2, 2], CA^2 = [1, 2, 1, 2]$ and the power allocations are $P^1$(in mw)$ = [10, 10, 10, 0], P^2$(in mw)$ = [0, 0, 10, 15]$  with allocated rates (in kbps) $[227.0, 111.0], [207.7, 201.3]$. The total sum power in the system is a little higher than in the above Pareto point but lower than those in the two CCEs above.\\
\begin{figure}
  \centering
  \includegraphics[scale=0.2]{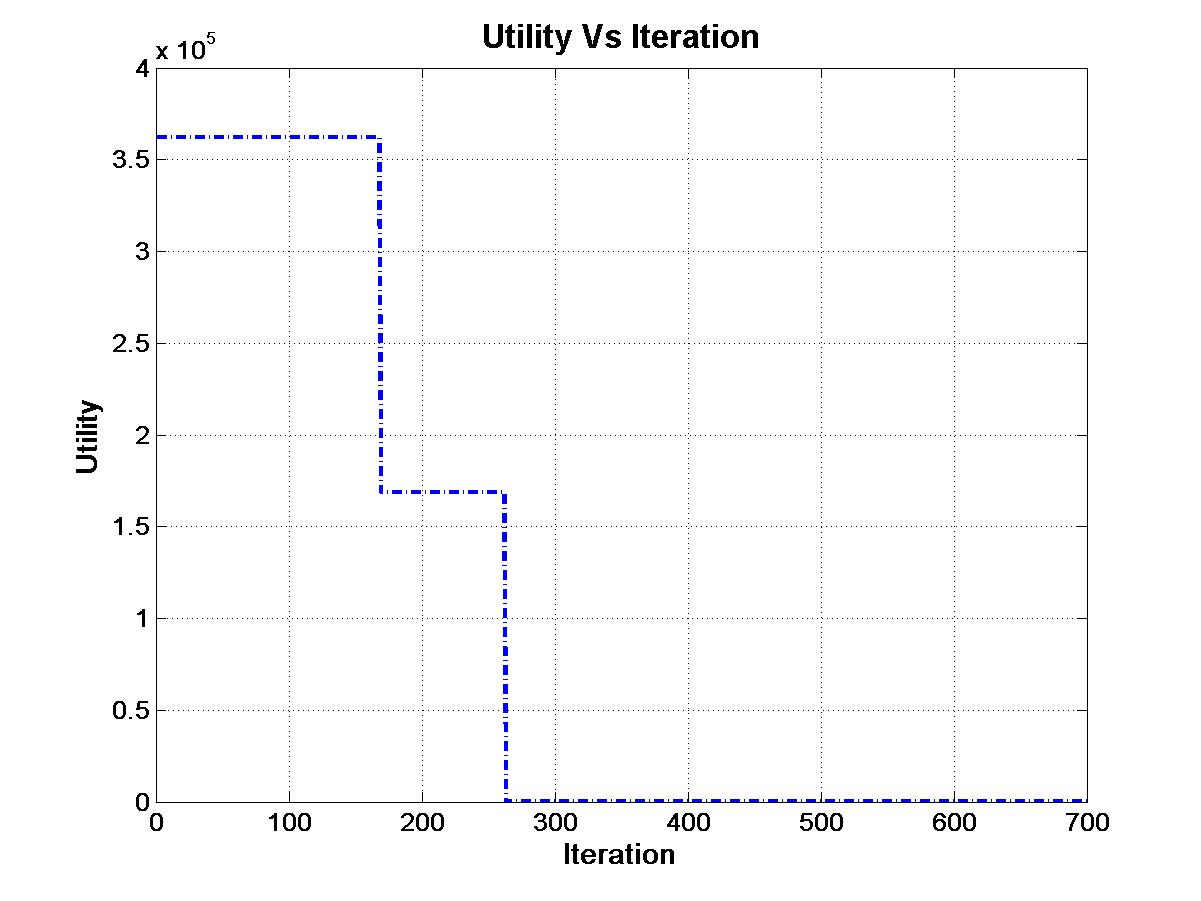}\\
  \vspace{-4mm}
  \caption{Example 2, Pareto point using Algorithm-3, eq (12).}
  %\vspace{-6mm}
\end{figure}
\textbf{Voice and Data Users}:

Next we consider a system with $2$ FCs, each with $2$ voice users and $1$ data user (for FC $1$, $\{2,3\}$ are voice users and $\{1\}$ is data user; for FC $2$, $\{1,3\}$ are voice users and $\{2\}$ is data user) and $4$ subchannels.
We generated the subchannel gain matrices $G^1, G^2, G^{1,2}, G^{2,1}$ as before.
The interference matrices (in mw) are \\\[I^1 = \left(
                                  \begin{array}{ccc}
                                      10 &   6 &    6\\
     3   &  9  &  12\\
     1   & 11  &  12 \\
     9   &  5  &   3\\
       \end{array}
                                \right), I^2 = \left(
                                  \begin{array}{ccc}
                                     10 &   7  &   5\\
     1   &  5  &   2\\
     7   & 10  &   8\\
     7   & 11  &   3
              \end{array}
       \right)\]\\
Maximum powers (mw) allocated to the subchannels are,
\[\bar{P}^1 = \left[\begin{array}{cccc}
   2.5,  &  1.3,  &  5.0,  &  1.6
          \end{array}\right],\]\vspace{-1cm}\\
\[\bar{P}^2 = \left[\begin{array}{cccc}
   2.1,  &  1.0,  &  3.6,  &  1.7
           \end{array}\right].\]
Rate requirements of the users, in kbps, are, \[\bar{R}^1 = \left[\begin{array}{ccc}
220, & 25, &17
                                             \end{array}\right], \bar{R}^2 = \left[\begin{array}{ccc}
                    15,&180,&30                    \end{array}\right].\]

Convergence of utilities for CCE is shown in Fig 12. At the equilibrium point, subchannel allocation for FCs are $CA^1 = [1, 2, 1, 3], CA^2 = [2, 3, 2, 1]$ and the allocated powers are $P^1$(in mw)$ = [2.5, 0.3, 5.0, 0.2], P^2$(in mw)$ = [2.1, 0.3, 3.6, 0.5]$ with allocated rates (in kbps) $[211.23, 25.06, 17.59], [15.19, 122.9, 35.6]$. The utilities are $\alpha_1=0.9601$ and $\alpha_2=0.6828$. The total allocated power are $8.0, 6.6$.
\begin{figure}
  \centering
  \includegraphics[scale=0.2]{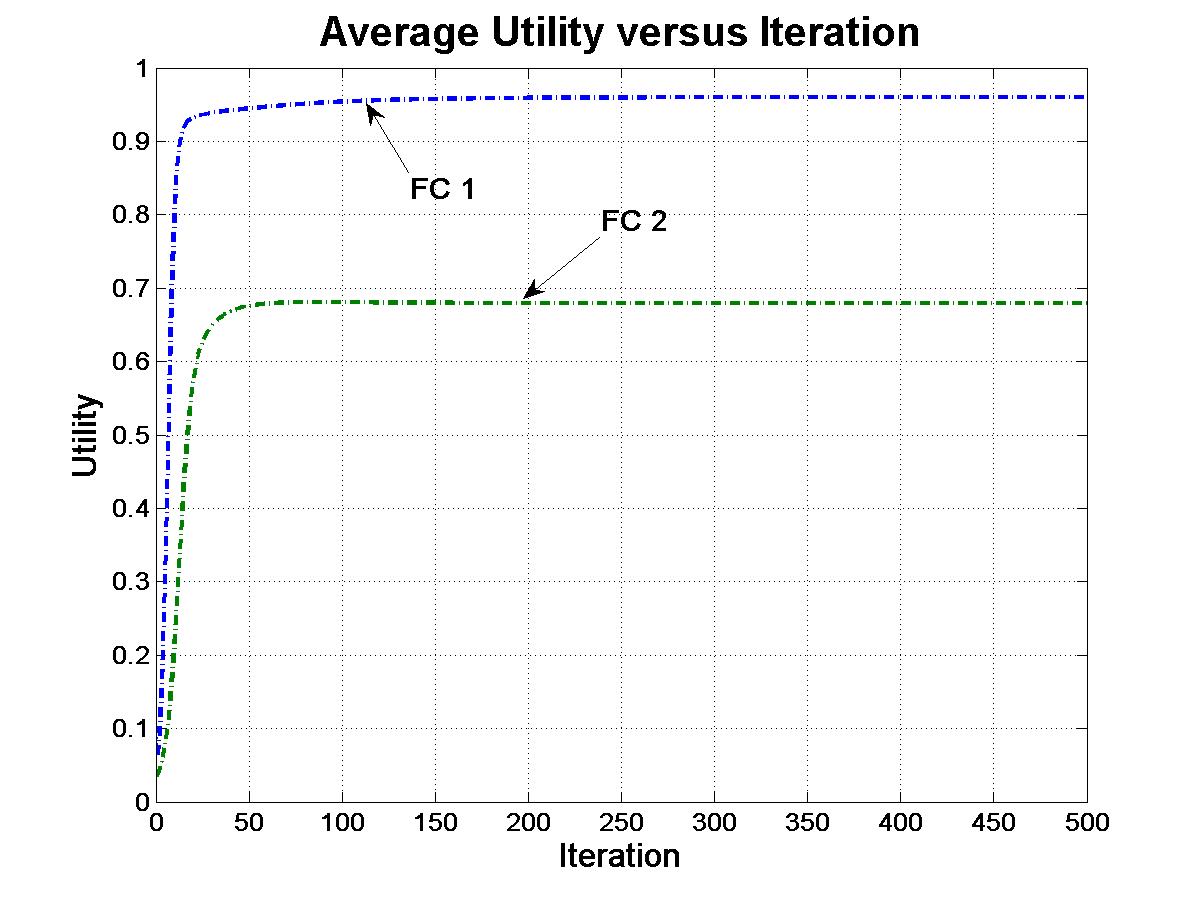}\\
  \vspace{-3mm}
  \caption{Example 3, CCE for voice and data user.}
  %\vspace{-6mm}
\end{figure}

Utilities of the Pareto points obtained for LHS of (22) using equation (14) and Algorithm 3 are shown in Fig 13. The subchannel allocation is $CA^1 = [1, 2, 1, 3], CA^2 = [2, 3, 2, 1]$ and the power allocations are $P^1$(in mw)$ = [1.7, 0.3, 5.0, 0.2], P^2$(in mw)$ = [2.1, 0.3, 3.6, 0.5]$  with allocated rates (in kbps) $[205.0, 25.04, 17.04], [15.5, 123.9, 35.6]$.  The utilities are $\alpha_1=0.9318$ and $\alpha_2=0.6883$.
\begin{figure}
  \centering
  \includegraphics[scale=0.2]{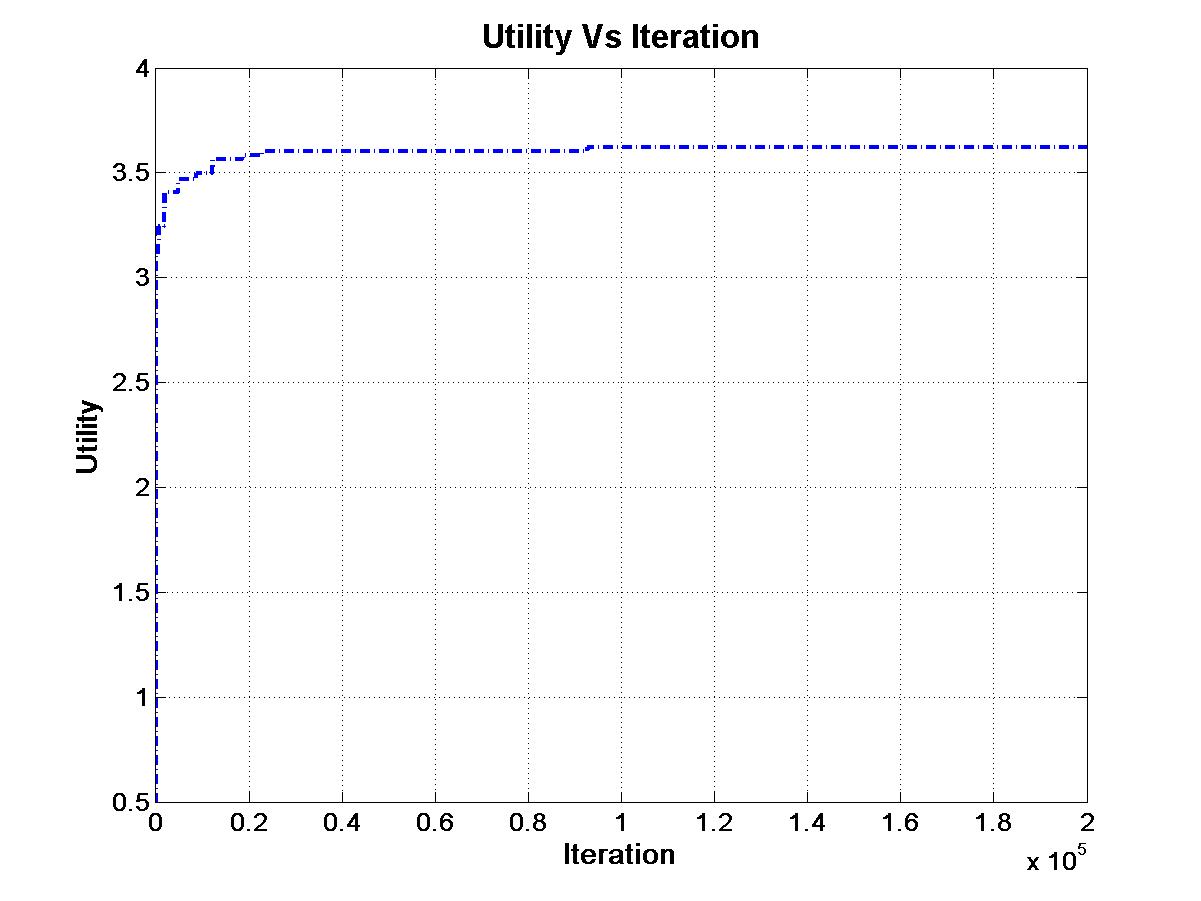}\\
  \vspace{-4mm}
  \caption{Example 3, Pareto point using Algorithm-3, eq (22).}
  %\vspace{-6mm}
\end{figure}

Utilities of the Pareto points obtained for (23) using equation (14) and Algorithm 3 are shown in Fig 14. The subchannel allocation is $CA^1 = [1, 2, 1, 3], CA^2 = [2, 3, 2, 1]$ and the power allocations are $P^1$(in mw)$ = [2.5, 0.3, 5.0, 0.2], P^2$(in mw)$ = [2.1, 0.3, 3.6, 0.5]$  with allocated rates (in kbps) $[211.2, 25.04, 17.06], [15.1, 122.2, 35.9]$. Also, $\alpha_1=0.9601, \alpha_2=0.6789$. Now the total power used is more than that of the above Pareto point.
\begin{figure}
  \centering
  \includegraphics[scale=0.2]{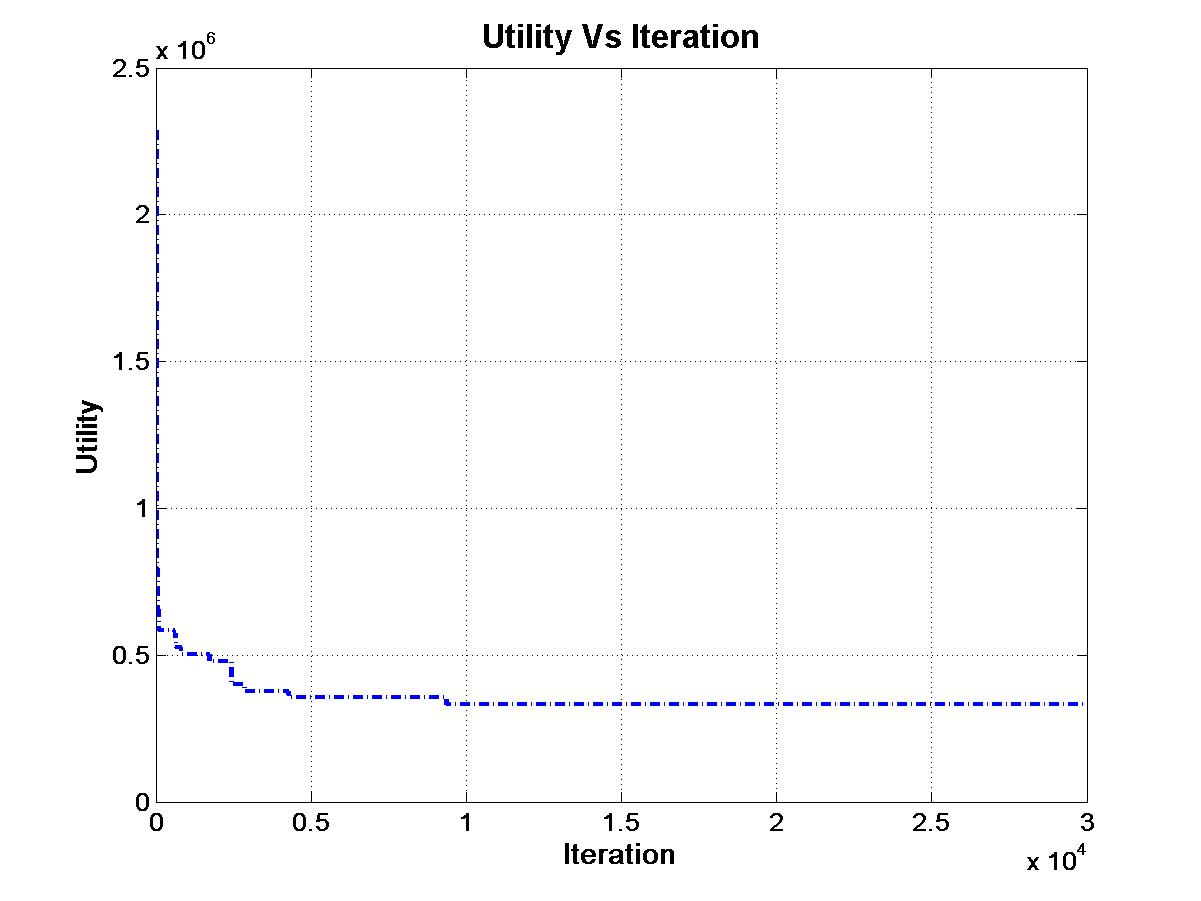}\\
  \vspace{-4mm}
  \caption{Example 3, Pareto point using Algorithm-3, eq (23).}
  %\vspace{-6mm}
\end{figure}

Utilities of the Nash Bargaining solution with $\boldsymbol{\hat{\alpha}}=\{1.4, 1.4\}$ obtained for LHS of (22) using equation (17) is shown in Fig 15. The subchannel allocation is $CA^1 = [1, 2, 1, 3], CA^2 = [2, 3, 2, 1]$ and the power allocations are $P^1$(in mw)$ = [2.5, 0.3, 5.0, 0.2], P^2$(in mw)$ = [2.1, 0.3, 3.6, 0.6]$  with allocated rates (in kbps) $[211.23, 25.06, 17.31], [15.8, 122.25, 35.62]$. It is observed that the rates allocated to the data users are almost same as that obtained via the Pareto optimal point obtained using LHS of (22). This solution does not improve fairness compared to the above CCE and Pareto points. Thus, to improve fairness, we changed $(\hat{\alpha}_1,\hat{\alpha}_2)$ to higher values. But it still did not improve the fairness. Taking $(\hat{\alpha}_1,\hat{\alpha}_2)$ larger either gives the same solution or makes it infeasible.
\begin{figure}
  \centering
  \includegraphics[scale=0.2]{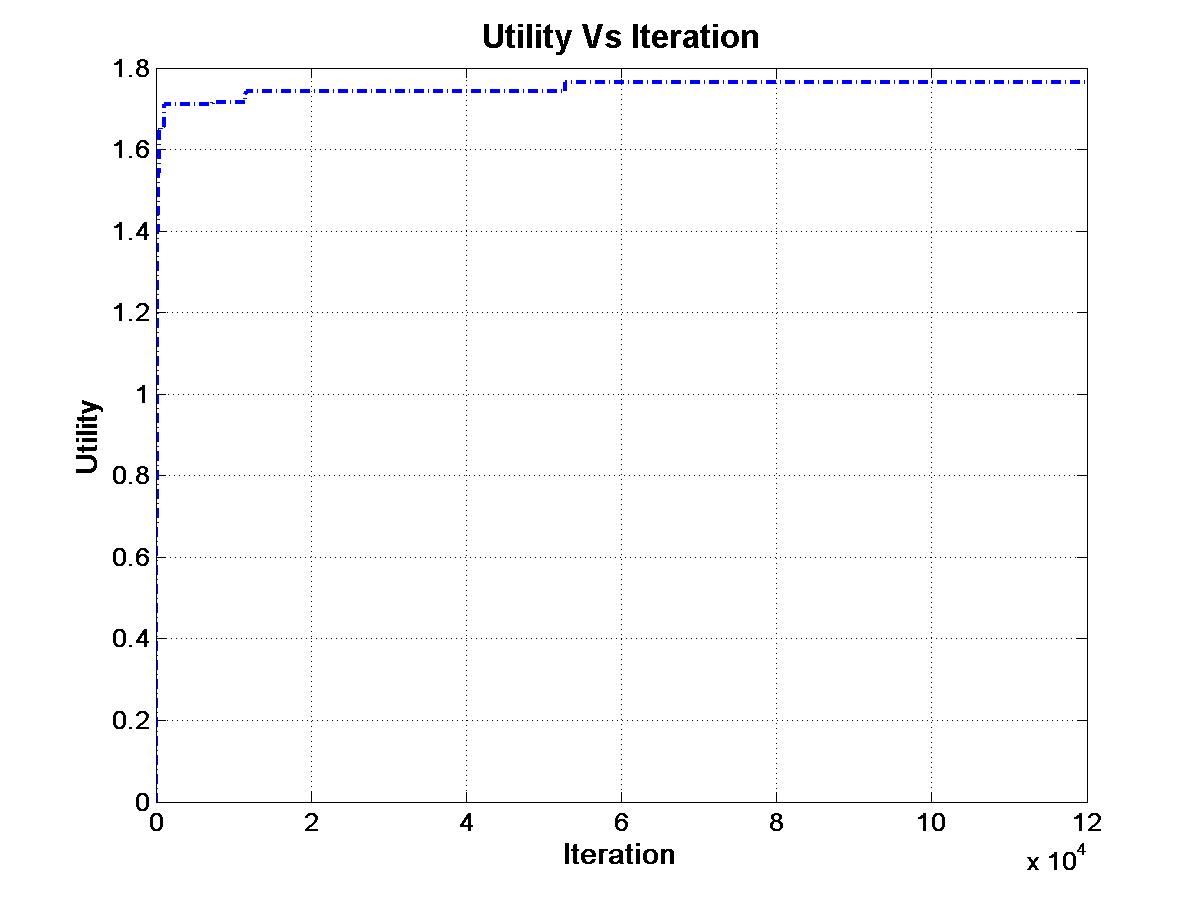}\\
  \vspace{-4mm}
  \caption{Example 3, Nash Bargaining solution using Algorithm-3, eq (22).}
  %\vspace{-6mm}
\end{figure}
\section{Conclusions}
We have considered a channel and power allocation problem in a multiple FC environment when there are multiple users in each FC requiring different minimum rates. The channels need to be shared by the different FCs such that they do not cause much interference to each other and to the MC users. We have formulated the problem in a non-cooperative game theory setup and provided low complexity distributed algorithms to obtain coarse correlated equilibria. We have also provided a distributed algorithm to obtain Pareto points and Nash bargaining solutions. For the Pareto points (also for the NB solution) we have also obtained a powerful low complexity heuristic. Finally we have extended these algorithms to a system with voice and data users, when the voice users are guaranteed their rates.
%\section{Acknowledgments}
%The authors would like to thank Dr. M.H.Kori, former technical director of Alcatel-Lucent for technical discussions.


\begin{thebibliography}{1}
\bibitem{abdel}
A Abdelnasser, E Hossain, D I Kim, ``Tier-Aware Resource allocation in OFDMA Microcell-Smallcell Networks", \emph{IEEE Transactions on Communications}, Vol. 63, No. 3, March, 2015.
\bibitem{barba}
S. Barbarossa, S. Sardellitti, A. Carfagna, P. Vechiarelli, ``Decentralized Interference Management in Femtocells: A Game theoretic Approach", \emph{IEEE Crowncom Proceedings, 2010}.
\bibitem{mehdi}
M Bennis, M Debbah, A Chair, ``On spectrum sharing with underlaid Femtocell Networks", \emph{IEEE, PIMRC-2010}.
\bibitem{vick1}
V. Chandrasekhar, J. Andrews, and A. Gatherer, ``Femtocell networks: a survey", \emph{Communicatins Magazine, IEEE}, vol.46, pp.59-67, September, 2008.
\bibitem{vikram}
V Chandrasekhar, J G. Andrews, Z Shen, T Muharemovic, A Gatherer, ``Distributed Power control in Femtocell-Underlay Cellular Networks", \emph{IEEE Globecomm, 2009}.
\bibitem{chow1}
P S.Chow, J M. Cioffi, J A.C. Bingham, ``A practical discrete multitone transceiver loading algorithm for data transmission over spectrally shaped channels", \emph{IEEE Trans Commu}, Vol. 43, No. 2/3/4, February/March/April, 1995.
\bibitem{chow}
M Z. Chowdhury, Y.M. Jang, Z.J. Hass, ``Network evolution and QoS provisioning for integrated Femtocell/Macrocell Netowrks", \emph{Int Journal of Wireless and Mobile Networks (IJWMN)}, Vol.2, No.3, August, 2010.
\bibitem{ref_d}
J Deng, R Zhang, L Song, Z Han, G Yang, B Jiao, ``Joint power control and subchannel allocation for OFDMA Femtocell networks
using distributed auction game", \emph{International Conference on Wireless Communications and Signal Processing}, 2012.
\bibitem{decom}
F Facchinei, V Piccialli, M Sciandrone, ``Decomposition Algorithms for generalized Potential Games", \emph{Computational Optimizaton and Applications}, Vol. 50, Issue. 2, pp. 237-262, October, 2011.
\bibitem{fuden}
D Fudenberg, J Tirole, ``Game Theory", \emph{Ane Books Pvt. Ltd, 2005}.
\bibitem{ghosh}
A Ghosh et al., ``Heterogenious Cellular Networks: From Theory to Practice", \emph{IEEE Communicatons Magazine}, June, 2012.
\bibitem{Han1}
Z Han, D Niyato, W Saad, T Basar and A Hj{\o}rungnes, ``Game Theory in Wireless and Communication Networks: Theory, Models, and Applications", Cambridge University Press, 2012.
\bibitem{hart2}
S Hart, A Mas-Colell, ``A simple adaptive procedure leading to correlated equilibrium", \emph{Econometrica}, Vol. 68, No. 5, pp.1127-1150, September, 2000.
\bibitem{Hasan}
Z Hasan, H Boostanimehr, V K. Bhargava, ``Green Cellular Networks: A Survey, Some Research Issues and challenges", \emph{IEEE Communications, Surveys and Tutorials}, Vol. 13, No. 4, pp. 524-540, Fourth Quarter, 2011.
\bibitem{hong}
E J Hong, S Y Yun, D-H Cho, ``Decentralized Power control scheme in Femtocell networks: A Game theoretic Approach",\emph{IEEE International symposium on PIMRC-2009}.
%\bibitem{ref_c}
%J. Y. Kim, D-H Cho, ``A joint power and subchannel allocation scheme maximizing system capacity in indoor dense Femtocell downlink systems", \emph{IEEE PIMRC 2009}.
%\bibitem{Parag}
%P Kulkarni, W H Chin, T Farnham, ``Radio Resource Management Considerations for LTE Femtocells", \emph{ACM SIGCOM Computer Communication Review}, Volume 40, Number 1, pp. 26-30, January 2010.
\bibitem{Lasaulce}
S Lasaulce, H Tembine, ``Game Theory and Learning for Wireless Networks: Fundamentals and Applicatons", \emph{Elsevier Ltd.}, 2011.
\bibitem{ref_a}
L B Le, D Niyato, E Hossain, D I Kim, D T Hoang, ``QoS-Aware and energy efficient resource management in OFDMA Femtocells", \emph{IEEE Transactions on
Wireless Communications}, Vol. 12, No. 1, pp. 180-194, January, 2013.
\bibitem{yloong}
Y L Lee, J Loo, T C Chuah, ``Dynamic Resource Management for LTE based Hybrid access Femtocell Systems", \emph{IEEE Systems Journal}, Vol. PP, Issue. 99, pp. 1-12, June, 2016.
\bibitem{yllee}
Y L Lee, J Loo, T C Chuah, A A. El-Saleh, ``Multi-objective Resource allocation for LTE/LTE-A Femtocell/HeNB Networks using Ant colony Optimization", \emph{Wireless Personal Communications}, Springer Journal, pp. 1-22, Auguest, 2016.
\bibitem{ref_90}
S-Y Lian, Y-Y Lin, K-C Chen, ``Cognitive and Game theoritical Radio Resource Management for Autonomous Femtocells with QoS guarantees", \emph{IEEE Transcations on Wireless Communicatons}, Vol. 10, No. 7, pp. 2196-2206, July, 2011.
%\bibitem{David}
%D L\'opez-p\'erez, A Valcarce, G de la Roche and J Zhang, ``OFDMA Femtocells: A Roadmap on Interference Avoidance", \emph{IEEE Communication Magazine}, Vol. 47, pp. 41-48, September 2009.
\bibitem{roger}
R B. Myerson, ``Game Theory: Analysis of Conflict", Harvard University Press, October, 1997.
\bibitem{Miettinen}
K M. Miettinen, ``Nonlinear Multiobjective Optimization", \emph{Kluwer Academic Publishers}, 1999.
\bibitem{shapely}
D. Monderer, L S. Shapley, ``Potential Games", Games and Economic Behavior 14, 124-143 (1996).
\bibitem{neel}
J O'D Neel, ``Analysis and Design of Cognitive Radio Networks and Distributed Radio Resource Management Algorithms", PhD Dissertation, Virginia Polytechnic Institute and State University, Blacksburg, VA, 2006.
\bibitem{trong}
D T Ngo, L B Le, T L-Ngoc, ``Distributed Pareto-Optimal Power control for Utility maximization in Femtocell Networks", \emph{IEEE Trans. on Wireless Communications}, Vol. 11, No. 10, October, 2012.
\bibitem{rao}
J B. Rao, A O. Fapojuwo, ``A survey of Energy Efficient Resource Managemnet Techniques for Multicell Cellular Networks", \emph{IEEE Communicatins Surveys and Tutorials}, Vol. 16, No. 1, First Quarter, 2014.
\bibitem{rough}
T. Roughgarden, ``Lecture Notes on Algorithmic Game Theory", from the CS364A course (Fall 2013).
\bibitem{garden}
T. Roughgarden, ``Intrinsic Robustness of the Price of Anarchy", Proceedings of the 41st ACM Symposium on Theory of Computing (STOC), pp. 513–522, 2009.
\bibitem{uday1}
V U Sankar and V. Sharma, ``QoS Provisioning for Multiple Femtocells via Game Theory", \emph{Elsevier Journal of Computer Networks}, Vol 102, pp.70-82, 2016.
\bibitem{palomar}
G. Scutari, S. Barbarossa, D. P. Palomar, ``Potential Games: A framework for vector power control problems with coupled constraints", \emph{IEEE Conf. ICASSP 2006 Proceedings}.
\bibitem{shakir}
M Z Shakir et al., ``Green Heterogenious Small-cell Networks: Toward reducing the $CO_2$ emmissions of Mobile Communicatons Industry using Uplink Power Adaptation", \emph{IEEE Communications Magazine}, June, 2013.
\bibitem{noah}
N D. Stein, ``Characterization and computation of equilibria in infinite games", Masters thesis, MIT, June, 2007.
\bibitem{gills}
G Stoltz, G Lugosi, ``Learning correlated equilibria in games with compact sets of strategies",  \emph{Games and Economic Behavior}, Vol.59, pp. 187-208, April, 2007.
%\bibitem{wang}
%Y wang, M Quin, X Han, Y Zhow, J Shi, ``Game-Theoritic Power control for Interference Mitigation in Two-tier Smallcell Networks", \emph{IEEE $79^{th}$ Vehicular Technology Conforence}, pp. 1-5, May, 2014.
\bibitem{jxu}
J Xu, J Wang, Y Zhu, Y Yang, X Zheng, S Wang, L Liu, K Horneman, Y Teng, ``Cooperative distributed optimization for the Hyper-Dense Smallcell deployment", \emph{IEEE Communications Magazine}, May, 2014.
\bibitem{zahir}
T Zahir, K Arshad, A Nakata, K Moessner, ``Interference Management in Femtocells", \emph{IEEE Communicatons Surveys and Tutorials}, Vol. 15, No. 1, First Quarter, 2013.
\bibitem{ref_b}
H Zhang, W Zheng, X Chu, ``Joint Subchannel and Power allocation in interfrence limited OFDMA Femtocells with Heterogeneous QoS
guarantee", \emph{IEEE Globecom}, 2012.
\bibitem{ref6}
3GPP TR 36.921, 36.922: ``FDD, TDD Home eNodeB (HeNB) Radio Frequency (RF) requirements analysis".
\end{thebibliography}
\end{document}